# Hybrid Analog-Digital Channel Estimation and Beamforming: Training-Throughput Tradeoff (Draft with more Results and Details)

Tadilo Endeshaw Bogale, *Member, IEEE*, Long Bao Le, *Senior Member, IEEE* and

Xianbin Wang, *Senior Member, IEEE*

### Abstract

This paper designs hybrid analog-digital channel estimation and beamforming techniques for multiuser massive multiple input multiple output (MIMO) systems with limited number of radio frequency (RF) chains. For these systems, first we design novel minimum mean squared error (MMSE) hybrid analog-digital channel estimator by considering both perfect and imperfect channel covariance matrix knowledge cases. Then, we utilize the estimated channels to enable beamforming for data transmission. When the channel covariance matrices of all user equipments (UEs) are known perfectly, we show that there is a tradeoff between the training duration and throughput. Specifically, we exploit that the optimal training duration that maximizes the throughput depends on the covariance matrices of all UEs, number of RF chains and channel coherence time ($T_c$). We also show that the training time optimization problem can be formulated as a concave maximization problem for some system parameter settings where its global optimal solution is obtained efficiently using existing tools. In particular, when the base station equipped with 64 antennas and 1 RF chain is serving one single antenna UE, $T_c = 128$ symbol periods ($T_s$) and signal to noise ratio of 10dB, we have found that the optimal training durations are $4T_s$ and $20T_s$

Tadilo Endeshaw Bogale and Long Bao Le are with the Institute National de la Recherche Scientifique (INRS), Université du Québec, Montréal, QC, Canada, and Xianbin Wang is with the University of Western Ontario, London, ON, Canada. Email: {tadilo.bogale, long.le}@emt.inrs.ca and xianbin.wang@uwo.ca

for highly correlated and uncorrelated Rayleigh fading channel coefficients, respectively. The analytical expressions are validated by performing extensive Monte Carlo simulations.

## I. Introduction

Massive multiple input multiple output (MIMO) technology is one of the promising means for achieving the extremely high energy and spectrum efficiency requirements of the future 5G networks [1]. To exploit the full potential of a massive MIMO system, one can leverage the conventional digital architecture where all the signal processing is performed at the baseband frequency. In a digital architecture, a complete radio frequency (RF) chain is required for each antenna element at the transmitter and receiver, including low-noise amplifier, down-converter, digital to analog converter (DAC), analog to digital converter (ADC) and so on [2], [3]. Thus, when the number of base station (BS) antennas is very large (i.e., massive MIMO regime)[1], the high cost and power consumptions of mixed signal components, like high-resolution ADCs and DACs, makes it difficult to dedicate a separate RF chain for each antenna [5]–[10]. For these reasons, deploying massive MIMO systems with limited number of RF chains and hybrid analog-digital architecture has recently received significant attention. In this architecture, the digital signal processing can be realized at the baseband frequency using microprocessors whereas, the analog signal processing can be enabled at the RF frequency by employing low cost phase shifters [11]–[13].

Hybrid architecture can be deployed for single user and multiuser massive MIMO systems. The key difference between the analog and digital components of the hybrid architecture is that, as the analog part is realized using phase shifters (i.e., to reduce cost), each having a constant modulus, it is more constrained than that of the digital architecture. Furthermore, in the multiuser system, the digital component can be designed for each UE independently (i.e., it can be unique to each UE) whereas, the analog component must be the same for all UEs (i.e., the analog part cannot be altered adaptively for each UE). In [14]–[16], a hybrid architecture is suggested for single user massive MIMO systems where matching pursuit algorithm is utilized for hybrid bemforming [2]. In [17], a low complexity codebook based RF beamforming based on multi-level RF beamforming and level-adaptive antenna selection is considered. In [18], a hybrid beamforming design utilizing interleaved and side-by-side sub-arrays is proposed [19]. This design is used for adaptive angle of arrival (AoA) estimation and beamformings by utilizing two algorithms; differential beam tracking and beam search.

---

[1]To the best of our knowledge, there is no strict definition on how large the number of antennas should be to be called as massive MIMO. However, in a multiuser setup, the term "massive MIMO" is used to reflect that the number of BS antennas is much larger than the number of user equipments (UEs) [4].

In [20], hybrid precoding scheme for multiuser massive MIMO systems is considered. The paper employs the zero forcing (ZF) approach where it is designed to maximize the sum rate of all users. In [6], a beam alignment technique using adaptive subspace sampling and hierarchical beam codebooks is proposed for millimeter wave (mmWave) cellular networks. In [21], a beam domain reference signal design for downlink channel with hybrid beamforming architecture is proposed to maximize the gain in a certain direction around the main beam. Hybrid analog-digital beamforming is proposed in [22] for downlink multiuser massive MIMO systems where total sum rate maximization problem is considered. In [23], hybrid beamforming design for multiuser setup with frequency selective channels is considered. This paper also discusses the required number of RF chains and phase shifters such that the digital and hybrid beamforming designs achieve the same performance. In [24], joint channel estimation and beamforming design is considered for single user mmWave MIMO systems. This paper employs a codebook approach to design its trainings for uniform linear array (ULA) channel models.

Channel state information (CSI) acquisition is an important aspect of a (massive) MIMO system. In general, a wireless channel has a nonzero coherence time where the channel is treated as almost constant. In a typical setup, increasing the training duration improves the channel estimation quality at the expense of reduced overall system throughput when the channel coherence time is fixed. Hence, there is a tradeoff between the training duration and system throughput. In most practical cases, the CSI needs to be learned to maximize the system throughput. This motivates us to study the training-throughput tradeoff for massive MIMO systems with limited RF chains while leveraging the hybrid architecture. The current paper particularly aims at determining the optimal training duration such that the system throughput is maximized. As can be understood from the above discussions, there are a number of research works utilizing hybrid architectures. However, none of the aforementioned works consider the problems studied in the current paper. For example, the works of [14]–[16], [20], [22], [23] focus on the design of hybrid beamforming under the assumption of perfect CSI. Furthermore, the works of [18], [24] consider both channel estimation and beamforming with hybrid architecture without taking into account the channel coherence time (i.e., the channel estimation duration derived by the aforementioned works may not necessarily maximize the system throughput).

Given these discussions, however, the study of training-throughput tradeoff has been conducted in the conventional MIMO systems where the number of RF chains is the same as that of antennas. In [25], the training scheme for single user block fading channel under low signal to noise ratio (SNR) regions is considered [26]. For the Rayleigh fading channel, the paper addresses on how long the channel coherence time be such that the lower bound mutual information behaves closer to the capacity obtained when perfect CSI is available at the receiver. In [27], the ergodic capacity of ultra wideband communications having sparse multipath channels for single user system at low SNR regions is examined. In [28], the structure of the optimal capacity achieving input matrix is derived. This paper also shows that there is no capacity gain by deploying the number of transmit antennas to be more than the coherence interval (in symbols) of the channel. In [29], the noncoherent MIMO capacity in the high SNR regime is examined where it is shown that the number of transmit antennas required need not be more than half of the coherence interval (in symbols) in this regime.

In [30], the authors consider multiple antenna communications over a wideband noncoherent Rayleigh block fading channel. This paper examines the capacity of a MIMO system with an average power constraint, and considers its relationship with the channel coherence time, number of transmit and receive antennas and SNR. The relation between wideband capacity and channel sparsity is studied in [31]. The work of [32] computes the optimal number of UEs that can be served in a given channel coherence time. From these explanations, we can understand that the works of [25]–[28], [30], [31] focus on the capacity characterization of MIMO systems for different SNR regions and channel characteristics. On the other hand, [32] tries to determine the optimal number of UEs for the given channel coherence time. However, in the current paper, we have examined the optimal training duration using hybrid architecture for fixed number of UEs and channel coherence time. We would like to emphasize here that one may still think to directly use the solution derived in [32] for the setup of the current paper. However, as will be shown in Section VI-A and demonstrated in the simulation section, such approach leads to the sub-optimal solution, and the throughput gap is significant in some parameter settings.

The training-throughput tradeoff can be studied by taking into account different settings and assumptions. As massive MIMO systems have significant interest at millimeter wave frequencies [6], [14], [21], [33], one may be interested in examining the training-throughput tradeoff for these frequencies only. For instance, by exploiting the characteristics of mmWave channels which are often specular and of very low rank. However, massive MIMO systems still have practical interest at microwave frequency bands [1]. For this reason, we examine the training-throughput tradeoff for multiuser massive MIMO systems with limited RF chains where the channel of each UE has arbitrary rank. It is assumed that the number of UEs are fixed, the BS is equipped with a massive antenna array and each UE is equipped with single antenna, with flat fading channels and TDD based channel estimation. Under these assumptions, first, we design novel TDD minimum mean squared error (MMSE) based hybrid analog-digital channel estimator by considering both perfect and imperfect channel covariance matrix cases. Then, we utilize the estimated channels to exploit beamforming during data transmission. Finally, we study the training-throughput tradeoff for the scenario where the channel covariance matrices of all UEs are known perfectly. The extension of the proposed designs for the case where each UE is equipped with multiple antennas, and experiences a frequency selective channel has also been discussed briefly. The main contributions of the paper are summarized as

follows:

- First, we propose channel estimation for multiuser massive MIMO system with perfect and imperfect channel covariance matrices by assuming that the number of RF chains is the same as that of antennas. When the covariance matrix is not known perfectly, the global optimal solution of our channel estimator is obtained by utilizing convex optimization approach. Second, we exploit these estimators to design the hybrid analog-digital channel estimator and beamformings for the scenario where the number of RF chains is less than that of antennas. As will be clear later in Section II, unlike the approach of [24], the proposed hybrid channel estimator and digital beamforming design can be applied for any channel model.

- When the covariance matrices of all UEs are known perfectly, we study the training-throughput tradeoff for the multiuser massive MIMO systems having limited number of RF chains. We exploit the fact that the optimal training duration that maximizes the overall network throughput depends on the operating SNR, available number of RF chains, channel coherence time ($T_c$) and covariance matrices of all UEs. Specifically, we show that the training time optimization problem can be formulated as a concave maximization problem for some system parameter settings where its global optimal solution can be obtained efficiently using existing tools.

- We have validated the analytical expressions by performing numerical and extensive Monte Carlo simulations. In particular, when the base station equipped with $64$ antennas and $1$ RF chain is serving one single antenna UE, and SNR=10dB, $T_c = 128$ symbol periods ($T_s$), we have found that the optimal training durations are $4T_s$ and $20T_s$ for highly correlated and uncorrelated Rayleigh fading channel coefficients, respectively.

This paper is organized as follows: Section II introduces the system model and problem statement. Section III - V discusses the proposed channel estimation and transmission approaches when the number of RF chains is the same as and less than that of antennas. Numerical and computer simulation results are presented in Sections VI-A and VI-B. Finally, conclusions are drawn in Section VIII.

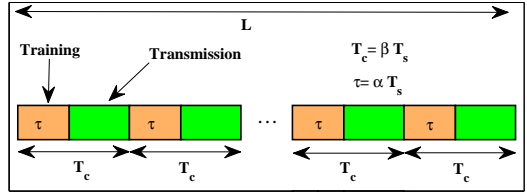

Fig. 1.   The considered channel estimation and transmission frame structure.

*Notations:* In this paper, upper/lower-case boldface letters denote matrices/column vectors. $\mathbf{X}_{(i,j)}$, $\mathrm{tr}(\mathbf{X})$, $\mathbf{X}^H$ and $\mathrm{E}(\mathbf{X})$ denote the $(i,j)$th element, trace, conjugate transpose and expected value of $\mathbf{X}$, respectively. We define $\|\mathbf{X}\|_F^2$ as $\mathrm{tr}\{\mathbf{X}^H\mathbf{X}\}$ and $\|\mathbf{X}\|_2$ as the square root of the maximum eigenvalue of $\mathbf{X}^H\mathbf{X}$, $\mathbf{I}$ is an identity matrix with appropriate size, $\mathcal{C}^{M \times M}$ and $\mathcal{R}^{M \times M}$ represent spaces of $M \times M$ matrices with complex and real entries, respectively. The next integer greater than or equal to $x$, optimal $x$ and $\max(x, 0)$ are denoted by $\lceil x \rceil$, $x^\star$ and $[x]_+$, respectively. The acronyms $\mathrm{blkdiag}(.)$, $\mathrm{s.t}$, $\mathrm{rem}$ and i.i.d denote "block diagonal", "subject to", "remainder" and "independent and identically distributed", respectively.

## II. System model and problem statement

This section discusses the system model and problem statement. We consider a multiuser system where a BS having $N$ antennas and $N_{RF}$ RF chains is serving $K$ decentralized single antenna UEs. A TDD based channel estimation is adopted where the channels of all UEs are estimated at the BS in the uplink channel, and are utilized for both the uplink and downlink data transmission phases. Under such settings, one can have a channel estimation and data transmission frame structure as shown in Fig. 1.

As we can see from this figure, one block incorporates $L$ frames for channel estimation and data transmission, where the duration of each frame is the same as the coherence time of the channel $T_c = \beta T_s$. In each frame, $\alpha T_s$ is used for channel estimation and the remaining $(\beta - \alpha)T_s$ is used for data transmission. In a typical massive MIMO system, the transmitted signal is received by an antenna array each described by the number of scatterers, AoA, angle of departure (AoD), delay, and fast fading parameter. Unlike the fast fading, the angles and delays are almost stationary over a considerable number of frames [34], [35].

This motivates us to assume that the $k$th UE has a constant channel covariance matrix $\mathbf{R}_k$, which does not depend on the fast fading component, in the $L$ frames of Fig. 1. Under this assumption, the channel between the $k$th UE and BS in each frame can be expressed as

$$\mathbf{h}_k = \mathbf{R}_k^{1/2}\tilde{\mathbf{h}}_k$$

where the entries of $\tilde{\mathbf{h}}_k \in \mathcal{C}^{N \times 1}$ are modeled as i.i.d zero mean circularly symmetric complex Gaussian (ZMCSCG) random variables each with unit variance, and $\mathbf{R}_k \in \mathcal{C}^{N \times N}$ is a positive semi-definite matrix.

Note that mmW channels are expected to be specular, consequently $\mathbf{R}_k$ will tend to be low rank. However, at microwave frequency bands, the rank of $\mathbf{R}_k$ will be likely high. Nevertheless, as the current paper does not assume any special structure on the channel, the above statistical channel model can be utilized both for mmWave and microwave frequency band applications. For microwave frequency regions, the assumption of perfectly (imperfectly) known $\mathbf{R}_k$ is commonly adopted in the existing literature. The Doppler shift scales linearly with frequency, thus, the coherence time of mmWave bands is an order of magnitude lower than that of comparable microwave bands. In a typical setting, for example, one may experience channel coherence times of $500\mu s$ and $35\mu s$ when the system is deployed at the carrier frequencies 2GHz and 28GHz, respectively [36], [37]. Furthermore, the fading channel statistics becomes wide-sense stationary (WSS) when the scattering geometry relative to a given user remains unchanged/varying slowly [38]. This validates that the covariance matrices of mmWave channels will also be almost constant for a considerable symbol duration [38], [39]. On the other hand, different channel covariance matrix estimation methods are recently proposed for millimeter wave channels in [40] by utilizing hybrid analog-digital architecture. These motivate us to consider the following objectives:

**Objectives of this paper**:

1) The first objective of this work is to design the channel estimation algorithm by assuming that $\mathbf{R}_k$ is known perfectly and imperfectly, and $N_{RF} < N$.

2) The ultimate goal of the channel estimation step is to enable reliable data transmission on $(\beta - \alpha)T_s$ durations. For this reason, the second objective of this work aims on the design of $\alpha$ such that the average throughput of each frame is maximized for the given $T_c$, $\beta$ and $N_{RF}$. This problem is mathematically formulated as[2]

$$\max_{\alpha} \ Th \tag{1}$$

where $Th$ is the total throughput obtained in each frame. To the best of our knowledge, when $\mathbf{R}_k$ is known imperfectly, the exact throughput expression is not known, and its computation requires considering significant bounds and expectations [43]. Therefore, we examine (1) and study the training-throughput tradeoff when $N_{RF} < N$ and perfect $\mathbf{R}_k$ scenario only[3].

Note that in some practical setup, the total energy available for both training and data transmission tasks remain fixed on the basis of the overal SNR level. For such a case, the training throughput tradeoff problem turns out to be on how to split the available energy into the training and data transmission phases. However, the training throughput tradeoff studied in the current paper considers different devices. Specifically, the training signal is transmitted from the UEs only (i.e., no training data is transmitted from the BS). And, it is not clear how one can assume fixed total energy that will be shared by the training and data communication phases for our TDD based channel estimation, and data transmission in both the uplink and dowlink channels. For these reasons, we have tried to compute the optimal training durations (in terms of symbol period) for maximizing the throughput while keeping the total coherence time and SNR levels constant (i.e., to solve (1)). Nevertheless, one can still utilize the idea of fixed total energy by considering both training and data transmission tasks in the uplink channel only where these tasks are performed by the UEs (or in a frequency division duplex (FDD) system, in general, where training and data signals are transmitted from the same source). And, since energy and SNR are interrelated quantitative terms, we believe that considering the training-throughput problem in terms of SNR (energy) will lead to the same result.

---

[2]Note that the current paper assumes that the channel covariance matrices of all UEs are available at the BS. Furthermore, the coherence time of the considered channel is known a priori. Although recent results at mmWave frequency bands support these assumptions for some parameter settings (see for example, [36], [37], [41], [42]), further measurement results are required to validate the considered assumptions for all practically relevant deployment scenarios.

[3]Studying the training-throughput tradeoff for imperfect $\mathbf{R}_k$ is left for future research.

## III. Channel Estimation when $N_{RF} = N$

In this section, we provide the proposed channel estimation approaches for the case where $N_{RF} = N$. For some environments, the covariance matrix of each UE is known a priori. For example, an i.i.d Rayleigh fading channel is commonly adopted in an urban environment (i.e., $\mathbf{R}_k = \mathbf{I}$); an exponential correlation model with correlation coefficient $\rho_k$ is also used to model the covariance matrix of each UE [44]. Furthermore, in a ULA channel model with sectorized antenna and predefined array geometry, the covariance matrix of each UE can also be known [45], [46]. However, the covariance matrix of each UE may not be known for a general communication environment. In such a case, it may need to be estimated from the received signal which introduces some error. In addition, as the current paper assumes a hybrid architecture with $N_{RF} \ll N$ RF chains, it exploits less degrees of freedom (compared to the system equipped with full number of RF chains) to estimate the covariance matrices of all UEs reliably in a practically relevant block duration. Consequently, the effect of channel covariance matrix estimation error could be prominent. These motivate us to design channel estimators for the following two cases:

1) **Case I**: In this case, it is assumed that $\mathbf{R}_k$ is perfectly known a priori.

2) **Case II**: In this case, we assume that $\mathbf{R}_k^{1/2} = \hat{\mathbf{R}}_k^{1/2} + \mathbf{\Delta}_k$ with $\|\mathbf{\Delta}_k\|_2 \le \epsilon_k$, where $\hat{\mathbf{R}}_k^{1/2}$ is the estimate of $\mathbf{R}_k^{1/2}$ and $\epsilon_k$ is the maximum error bound[4].

### A. Channel Estimation for **Case I**

As explained in Section I, the channel estimation of a massive MIMO system is performed in the uplink channel. During the uplink training phase, the BS will receive the following signal at symbol period $j$.

$$\mathbf{y}_{tj} = \sum_{i=1}^{K} \mathbf{h}_i p_{ij}^* + \mathbf{n}_{tj} \tag{2}$$

where $K$ is the number of UEs, $p_{ij}^*$ is the pilot symbol of the $i$th UE at symbol time $j$, $\mathbf{n}_j \in \mathcal{C}^{N \times 1}$ is the received noise vector at symbol time $j$ and $\mathbf{h}_i \in \mathcal{C}^{N \times 1}$ is the channel between the $i$th UE and BS.

---

[4]We would like to mention here that such bounded error parameter estimation approach is termed as worst case design [47].

When $N_{RF} = N$ and the coherence time of the channel is $T_c$, channel estimation is performed just by employing $\tau = K$ symbol periods ($T_s$) and data transmission is performed on the remaining $T_c - \tau$ time durations [32], [48]. By spending $K$ training symbols, the BS receives the following signal

$$\mathbf{Y}_t = \mathbf{HP}^H + \mathbf{N}_t$$

where $\mathbf{H} = [\mathbf{h}_1, \mathbf{h}_2, \cdots, \mathbf{h}_K]$, $\mathbf{Y}_t = [\mathbf{y}_{t1}, \mathbf{y}_{t2}, \cdots, \mathbf{y}_{tK}]$, $\mathbf{P} = [\mathbf{p}_1, \mathbf{p}_2, \cdots, \mathbf{p}_K]$, $\mathbf{N}_t = [\mathbf{n}_{t1}, \mathbf{n}_{t2}, \cdots, \mathbf{n}_{tK}]$ is the additive noise which is modeled as $\mathbf{N}_t \sim \mathcal{CN}(\mathbf{0}, \sigma_t^2 \mathbf{I})$ and $\mathbf{p}_i = [p_{i1}, p_{i2}, \cdots, p_{iK}]^T$. For simplicity, the entries of $\mathbf{P}$ are selected from a constellation satisfying $\mathbf{P}^H \mathbf{P} = \mathbf{I}$. With this training, the channel estimation phase follows two steps. The first step is used to decouple the channels of each UE independently and the second step is used to get the estimated channel of each UE from the decoupled terms.

In the first step, the BS can decouple the channels of each UE by multiplying the overall received signal $\mathbf{Y}$ with $\mathbf{P}$, i.e.,

$$\mathbf{E} \triangleq \mathbf{Y}_t \mathbf{P} = \mathbf{H} + \mathbf{N}_t \mathbf{P} \Rightarrow \mathbf{e}_k = \mathbf{h}_k + \tilde{\mathbf{n}}_k, \tag{3}$$

where $\tilde{\mathbf{n}}_k = \mathbf{N}_t \mathbf{p}_k$. As we can see from (3), $\mathbf{e}_k$ does not contain $\mathbf{h}_i, i \neq k$. In the second step, the well known MMSE estimator is employed to estimate the $k$th channel from $\mathbf{e}_k$. This is performed by introducing an MMSE matrix $\mathbf{W}_k^H$ for the $k$th UE and expressing the estimated channel $\widehat{\mathbf{h}}_k$ as

$$\widehat{\mathbf{h}}_k = \mathbf{W}_k^H \mathbf{e}_k, \tag{4}$$

where $\mathbf{W}_k$ is designed such that the MSE between $\widehat{\mathbf{h}}_k$ and $\mathbf{h}_k$ is minimized as follows

$$\xi_k = \mathrm{tr}\{\mathrm{E}\{|\widehat{\mathbf{h}}_k - \mathbf{h}_k|^2\}\} = \mathrm{tr}\{\mathrm{E}\{|\mathbf{W}_k^H(\mathbf{h}_k + \tilde{\mathbf{n}}_k) - \mathbf{h}_k|^2\}\}$$

$$= \mathrm{tr}\{(\mathbf{W}_k^H - \mathbf{I}_N)\mathbf{R}_k(\mathbf{W}_k^H - \mathbf{I}_N)^H + \sigma_t^2 \mathbf{W}_k^H \mathbf{W}_k\}. \tag{5}$$

The optimal $\mathbf{W}_k$ is obtained from the gradient of $\xi_k$ as

$$\frac{\partial \xi_k}{\partial \mathbf{W}_k^H} = 0 \Rightarrow \mathbf{W}_k^\star = (\mathbf{R}_k + \sigma_t^2 \mathbf{I})^{-1} \mathbf{R}_k. \tag{6}$$

Note that when $N_{RF} = N$ and $\mathbf{R}_k$ is known perfectly, the above channel estimation approach is commonly adopted in the existing literature and can be considered as an existing approach [32], [48].

## B. *Channel Estimation for* **Case II**

In this subsection, we provide the proposed channel estimation approach when $\mathbf{R}_k$ is not known perfectly (i.e., **Case II**). As we can see from the above subsection, the channel estimation phase has two main steps when $\mathbf{R}_k$ is known perfectly. The first step (i.e., user decoupling) can still be used even if $\mathbf{R}_k$ is not known perfectly. However, the second step depends on the knowledge of $\mathbf{R}_k$. Thus, the remaining issue will be on how to design $\mathbf{W}_k$ such that $\xi_k$ of (5) is minimized under imperfect $\mathbf{R}_k$. To this end, we propose robust convex optimization approach as follows.

$$\min_{\mathbf{W}_k} \ \|(\mathbf{W}_k^H - \mathbf{I}_N)(\hat{\mathbf{R}}_k^{1/2} + \boldsymbol{\Delta}_k)\|_F^2 + \sigma_t^2 \mathrm{tr}\{\mathbf{W}_k^H \mathbf{W}_k\}, \quad \text{s.t } \|\boldsymbol{\Delta}_k\|_2 \leq \epsilon_k. \tag{7}$$

This problem can be equivalently expressed as [49]

$$\min_{\mathbf{W}_k, \mathbf{D}_k} \ \mathrm{tr}\{\mathbf{D}_k\} + \sigma_t^2 \mathrm{tr}\{\mathbf{W}_k^H \mathbf{W}_k\} \tag{8}$$

$$\text{s.t } \|\boldsymbol{\Delta}_k\|_2 \leq \epsilon_k, \quad (\mathbf{W}_k^H - \mathbf{I}_N)(\hat{\mathbf{R}}_k^{1/2} + \boldsymbol{\Delta}_k)(\hat{\mathbf{R}}_k^{1/2} + \boldsymbol{\Delta}_k)^H(\mathbf{W}_k - \mathbf{I}_N) \preceq \mathbf{D}_k, \ \mathbf{D}_k \succeq \mathbf{0}$$

where $\succeq \mathbf{0}$ denotes positive semi-definite. Using Schur Complement, the last inequality becomes [49]

$$\begin{bmatrix} \mathbf{I} & (\hat{\mathbf{R}}_k^{1/2} + \boldsymbol{\Delta}_k)^H(\mathbf{W}_k - \mathbf{I}_N) \\ (\mathbf{W}_k^H - \mathbf{I}_N)(\hat{\mathbf{R}}_k^{1/2} + \boldsymbol{\Delta}_k) & \mathbf{D}_k \end{bmatrix} \succeq \mathbf{0}.$$

This inequality can also be expressed as

$$\mathbf{A}_k \succeq \mathbf{B}_k^H \boldsymbol{\Delta}_k \mathbf{C}_k + (\mathbf{B}_k^H \boldsymbol{\Delta}_k \mathbf{C}_k)^H \tag{9}$$

where

$$\mathbf{A}_k = \begin{bmatrix} \mathbf{I} & \hat{\mathbf{R}}_k^{1/2}(\mathbf{W}_k - \mathbf{I}_N) \\ (\mathbf{W}_k^H - \mathbf{I}_N)\hat{\mathbf{R}}_k^{1/2} & \mathbf{D}_k \end{bmatrix}, \quad \mathbf{B}_k = [\mathbf{0}, \ (\mathbf{W}_k - \mathbf{I}_N)], \quad \mathbf{C}_k = \quad [-\mathbf{I}, \ \mathbf{0}]. \tag{10}$$

Thus, the last constraint of (8) can be replaced by (9). To solve (8), we consider *Lemma 1* below.

*Lemma 1*: Let $\mathbf{A}, \mathbf{B}$ and $\mathbf{C}$ are given matrices with $\mathbf{A} = \mathbf{A}^H$, the relation

$$\mathbf{A} \succeq \mathbf{B}^H \boldsymbol{\Delta} \mathbf{C} + (\mathbf{B}^H \boldsymbol{\Delta} \mathbf{C})^H, \ \boldsymbol{\Delta} : \|\boldsymbol{\Delta}\|_2 \leq \epsilon \tag{11}$$

exists if and only if there exists $\lambda \geq 0$ and

$$\begin{bmatrix} \mathbf{A} - \lambda \mathbf{C}^H \mathbf{C} & -\epsilon \mathbf{B}^H \\ -\epsilon \mathbf{B} & \lambda \mathbf{I} \end{bmatrix} \succeq \mathbf{0}.$$

*Proof:* See *Proposition 2* of [50]. ∎

Using *Lemma 1*, problem (8) can be reformulated as

$$\min_{\mathbf{W}_k, \mathbf{D}_k, \mathbf{M}_k, \lambda_k \geq 0} \mathrm{tr}\{\mathbf{D}_k\} + \sigma_t^2 \mathrm{tr}\{\mathbf{M_k}\}, \text{ s.t } \begin{bmatrix} \mathbf{A}_k - \lambda_k \mathbf{C}_k^H \mathbf{C}_k & -\epsilon_k \mathbf{B}_k^H \\ -\epsilon_k \mathbf{B}_k & \lambda_k \mathbf{I} \end{bmatrix} \succeq \mathbf{0}, \quad \begin{bmatrix} \mathbf{M}_k & \mathbf{W}_k^H \\ \mathbf{W}_k & \mathbf{I} \end{bmatrix} \succeq \mathbf{0} \quad (12)$$

where $\mathbf{A}_k, \mathbf{B}_k$ and $\mathbf{C}_k$ are as defined in (10). This problem is the well known semi-definite programming problem where its global optimal solution can be found by applying interior point methods [49].

We would like to mention here that problem (7) has "similar" mathematical structure to that of the worst case robust MIMO beamforming problem (see for example (39) of [47] and (25) of [51]). Thus, the approach used to solve (7) can be considered as the modified versions of the techniques used in [47], [51]. Furthermore, as will be detailed in the next subsection, we have exploited the solution of (7) to handle the scenario where the number of RF chains is smaller than those of antennas.

### C. Low Complexity Channel Estimator

When $\mathbf{R}_k$ is known perfectly, the complexity of (6) is the same as that of matrix inversion which is $O(N^3)$. On the other hand, under imperfect $\mathbf{R}_k$, (12) has $6N^2 + 1$ real optimization variables, and two semi-definite constraints with $6N$ and $4N$ dimensions. The worst case computational complexity of (12) in terms of number of iterations is upper bounded by $O(\sqrt{6N + 4N})$ where the complexity of each iteration is on the order of $O((6N^2 + 1)^2((6N)^2 + (4N)^2))$. Hence, the overall complexity of solving (12) is upper bounded by $O(\sqrt{10N}(6N^2 + 1)^2(52N^2)) \approx O(N^{6.5})$ [52]. Although these complexities have polynomial order, they may be expensive especially when $N$ is very large (i.e., in massive MIMO systems).

In a typical indoor massive MIMO (mmWave) communication, $\mathbf{R}_k$ will often tend to be a very low rank matrix. Thus, one approach of reducing the complexity of the channel estimators is to exploit the rank of $\mathbf{R}_k$ which is explained as follows. From the eigenvalue decomposition of $\mathbf{R}_k$, we will have $\mathbf{R}_k = \mathbf{Q}_k \boldsymbol{\Lambda}_k \mathbf{Q}^H$, where $\mathbf{Q}_k \in \mathcal{C}^{N \times L_k}$ with $L_k$ as the rank of $\mathbf{R}_k$, and we can reexpress $\mathbf{h}_k$ as

$$\mathbf{h}_k = \mathbf{Q}_k \bar{\tilde{\mathbf{h}}}_k, \Rightarrow \mathbf{Q}_k^H \mathbf{h}_k = \bar{\tilde{\mathbf{h}}}_k \tag{13}$$

where $\bar{\tilde{\mathbf{h}}}_k \in \mathcal{C}^{L_k \times 1} = \boldsymbol{\Lambda}_k^{1/2} \tilde{\mathbf{h}}_k$ is modeled $\mathcal{CN}(\mathbf{0}, \boldsymbol{\Lambda}_k)$. As we can see, $\mathbf{h}_k$ is the product of $\mathbf{Q}_k$ and $\bar{\tilde{\mathbf{h}}}_k$. The main idea of reducing the complexity of the channel estimators is to first estimate $\bar{\tilde{\mathbf{h}}}_k$ and then to multiply the estimated value just by $\mathbf{Q}_k$. Using (3) and (13), we can have

$$\mathbf{e}_k = \mathbf{Q}_k \bar{\tilde{\mathbf{h}}}_k + \tilde{\mathbf{n}}_k, \Rightarrow \mathbf{Q}_k^H \mathbf{e}_k = \bar{\tilde{\mathbf{h}}}_k + \mathbf{Q}_k^H \tilde{\mathbf{n}}_k.$$

By introducing $\tilde{\mathbf{W}}_k^H$, one can express the estimated $\bar{\tilde{\mathbf{h}}}_k$ as

$$\widehat{\bar{\tilde{\mathbf{h}}}}_k = \tilde{\mathbf{W}}_k^H \mathbf{Q}_k^H \mathbf{e}_k. \tag{14}$$

When $\mathbf{R}_k$ are known perfectly, $\tilde{\mathbf{W}}_k$ can be designed using the MMSE method like in (6). When $\mathbf{R}_k$ is not known perfectly and $\mathbf{e}_k$ is multiplied by $\mathbf{Q}_k^H$, the error covariance becomes $\mathbf{Q}_k^H \boldsymbol{\Delta}_k$. From the Cauchy-Schwarz inequality, we will have

$$\|\mathbf{Q}_k^H \boldsymbol{\Delta}_k\|_2 \leq \|\mathbf{Q}_k^H\|_2 \|\boldsymbol{\Delta}_k\|_2 = \|\boldsymbol{\Delta}_k\|_2 \leq \epsilon_k. \tag{15}$$

As the error bound is still maintained in such a case, the optimal $\tilde{\mathbf{W}}_k^H$ can be obtained similar to (12) when $\mathbf{R}_k$ is not known perfectly.

Once $\tilde{\mathbf{W}}_k$ is designed, the estimated channel can be computed as

$$\widehat{\mathbf{h}}_k = \mathbf{Q}_k \widehat{\bar{\tilde{\mathbf{h}}}}_k. \tag{16}$$

From the above explanation, we can understand that the complexity of the channel estimation (16) becomes $O(L_k^3)$ and $O(L_k^{6.5})$ for perfect and imperfect $\mathbf{R}_k$, respectively. And in practice as $L_k \ll N$, this channel estimation approach can reduce the complexity significantly.

Note that the channel estimators of this section are designed by assuming $\alpha = K$. However, in some cases, there could be an interest to design the channel estimator for more general training duration such as $\alpha = \tilde{\theta} K$, where $\tilde{\theta}$ is a positive integer. For such a case, the same pilot symbols as those of this section can be reused for $K$ consecutive symbol periods, and utilize linear combination approach. By doing so, one can obtain similar structure as $\mathbf{e}_k$ of (3) but the effective noise variance is reduced by a factor of $\tilde{\theta}$. This shows that the estimation techniques of this section can be applied straightforwardly for such cases.

**Design perspective**: As explained in the previous section, the ultimate goal of the channel estimator is to enable data transmission. In this regard, two design approaches are commonly adopted, one is by assuming perfect CSI and the other is by considering imperfect CSI. Perfect CSI can almost be achieved when $\mathbf{R}_k$ is known perfectly and $\sigma_t^2 \ll 1$ (i.e., at very high SNR). In all other cases, perfect CSI can never be achieved. Thus, in general, the transmission phase should also take into account the effect of imperfect CSI especially due to the imperfect $\mathbf{R}_k$. However, designing the transceivers by assuming imperfect CSI is not trivial and it requires a robust approach for each frame duration which is almost infeasible particularly for large antenna array systems (see for example [43], [47], [53], [54]). For this reason, the estimated channel obtained by utilizing the $\mathbf{W}_k$ of (6) and (12) or $\tilde{\mathbf{W}}_k$ of (14) is treated as perfect channel for data transmission phase[5]. As we can see from Fig. 1, the MMSE channel estimator $\mathbf{W}_k(\tilde{\mathbf{W}}_k)$ is designed only once in each block. This validates that the considered MMSE channel estimators have practical interest.

Once the channel estimation phase is accomplished, the next phase will be to perform data transmission. As discussed above, data transmission may take place in both the uplink and downlink channels. During data transmission phases, the BS exploits antenna arrays to design the appropriate beamforming matrices. In the current paper, we employ the well known ZF approach in both the downlink and uplink transmission phases. This is because ZF approach is optimal in a rich scattering environment and easy to implement for massive MIMO systems [4], [23]. When $\mathbf{R}_k$ is known perfectly, $N_{RF} = N$ and given $K$, the optimal solution of (1) is $\alpha = K$ [32].

In the following, we discuss the channel estimation and data transmission phases when $N_{RF} < N$. As will be clear in the sequel, the channel estimator of this section is exploited to design the channel estimator and study the training-throughput tradeoff when $N_{RF} < N$. For better flow of the text, we first

---

[5]We are particularly interested for high SNR regions $\sigma_t^2 \ll 1$.

start with single user case and then we proceed to multiuser.

## IV. Channel Estimation and Training-Throughput Tradeoff for Single User when

$$N_{RF} < N$$

This section discusses the channel estimation and studies the tradeoff between the training duration and throughput when $N_{RF} < N$ for the single user case.

### A. Channel Estimation

For the single user setup, the BS will receive the following signal in all antennas during the training duration (i.e., (2) with $K = 1$). Throughout this section, we have removed the subscript 1 for readability.

$$\mathbf{y} = \mathbf{h}p^* + \mathbf{n}$$

where $p$ is chosen as $|p|^2 = 1$. As mentioned in the introduction section, the hybrid analog-digital channel estimator consists of the analog and digital part, where the analog part is realized at the RF frequencies. Assume that we employ $\alpha$ trainings to perform channel estimation and $\bar{\mathbf{U}}_i \in \mathcal{C}^{N \times N_{RF}}$ is used as the analog channel estimator in the $i$th pilot symbol. With this matrix, the received samples after ADC in the $i$th pilot symbol can be expressed as

$$\tilde{\mathbf{y}}_i = \bar{\mathbf{U}}_i^H (\mathbf{h}p^* + \mathbf{n}_i). \tag{17}$$

By multiplying this signal with $p$, we will get

$$\tilde{\mathbf{h}}_i = p\tilde{\mathbf{y}}_i = \bar{\mathbf{U}}_i^H (\mathbf{h} + \mathbf{n}_i p). \tag{18}$$

Now if we employ $\alpha$ pilot symbols for training, we will have

$$\tilde{\mathbf{h}} = \bar{\mathbf{U}}^H \mathbf{h} + \bar{\mathbf{n}} \tag{19}$$

where $\tilde{\mathbf{h}} = [\tilde{\mathbf{h}}_1^T, \tilde{\mathbf{h}}_2^T, \cdots, \tilde{\mathbf{h}}_\alpha^T]^T$, $\bar{\mathbf{U}} = [\bar{\mathbf{U}}_1, \bar{\mathbf{U}}_2, \cdots, \bar{\mathbf{U}}_\alpha]$, $\bar{\mathbf{n}}_i = \bar{\mathbf{U}}_i^H \mathbf{n}_i p$ and $\bar{\mathbf{n}} = [\bar{\mathbf{n}}_1^T, \bar{\mathbf{n}}_2^T, \cdots, \bar{\mathbf{n}}_\alpha^T]^T$. Finally, we utilize the digital channel estimator $\bar{\mathbf{W}}$ to estimate $\mathbf{h}$ from $\tilde{\mathbf{h}}$ as follows

$$\bar{\mathbf{h}} = \bar{\mathbf{W}}^H \tilde{\mathbf{h}}. \tag{20}$$

By doing so, we will get the following MSE $\bar{\xi}$

$$\bar{\xi} = \text{tr}\{\text{E}\{|\bar{\mathbf{h}} - \mathbf{h}|^2\}\} = \text{tr}\{(\bar{\mathbf{W}}^H\bar{\mathbf{U}}^H - \mathbf{I}_N)\mathbf{R}(\bar{\mathbf{W}}^H\bar{\mathbf{U}}^H - \mathbf{I}_N)^H + \bar{\mathbf{W}}^H\mathbf{R}_{\bar{n}}\bar{\mathbf{W}}\} \tag{21}$$

where $\mathbf{R}$ is the covariance matrix of $\mathbf{h}$, $\mathbf{R}_{\bar{n}} = \text{E}\{\bar{\mathbf{n}}\bar{\mathbf{n}}^H\}$. Thus, when $\mathbf{R}$ is known perfectly, our objective will now be to design $\bar{\mathbf{W}}$ and $\bar{\mathbf{U}}$ to solve

$$\min_{\bar{\mathbf{W}},\bar{\mathbf{U}}} \|(\bar{\mathbf{W}}^H\bar{\mathbf{U}}^H - \mathbf{I}_N)\sqrt{\mathbf{R}}\|_F^2 + \bar{\mathbf{W}}^H\mathbf{R}_{\bar{n}}\bar{\mathbf{W}}. \tag{22}$$

And when $\mathbf{R}$ is not known perfectly, we will have the following problem

$$\min_{\bar{\mathbf{W}},\bar{\mathbf{U}}} \|(\bar{\mathbf{W}}^H\bar{\mathbf{U}}^H - \mathbf{I}_N)(\hat{\mathbf{R}}^{1/2} + \mathbf{\Delta})\|_F^2 + \bar{\mathbf{W}}^H\mathbf{R}_{\bar{n}}\bar{\mathbf{W}}, \quad \text{s.t } \|\mathbf{\Delta}\|_2 \leq \epsilon. \tag{23}$$

In general, the problems (22) and (23) are non-convex. However, for fixed $\bar{\mathbf{U}}(\bar{\mathbf{W}})$ these problems are convex. Therefore, one approach of solving such problems is to iteratively optimize one variable while keeping the other constant. In a massive MIMO system, as $N$ is typically large, solving these two problems iteratively is computationally expensive. For this reason, we preselect $\bar{\mathbf{U}}$ from the solution of Section III and then we optimize $\bar{\mathbf{W}}$ for fixed $\bar{\mathbf{U}}$. The question now is how do we select $\bar{\mathbf{U}}$?

In the following, we provide the proposed method to select $\bar{\mathbf{U}}$ when $\mathbf{R}$ is known perfectly[6]. By computing the singular value decomposition (SVD) of $\mathbf{W}^\star$ in (6), one can get

$$\mathbf{W}^\star = \mathbf{U}\mathbf{D}\mathbf{V}^H \tag{24}$$

where $\mathbf{D}$ is a diagonal matrix containing the non-zero singular values of $\mathbf{W}^\star$ arranged in decreasing order.

According to the detailed results discussed in [23], when each of the elements of $\bar{\mathbf{U}}$ has a modulus less than or equal to 2, this matrix can be implemented using phase shifters in two approaches. The first approach assumes the availability of sufficiently high resolution (or theoretically infinite) phase shifters (i.e., as in Section IV.A of [23]) whereas, the second approach assumes to utilize switches and constant phase shifters (i.e., the phases of each phase shifter is fixed a priori as in Section IV.B of [23]). Given the limitations of the current electronic technology, it is almost infeasible to have low cost and sufficiently high resolution phase shifters. For this reason, we believe that the second implementation approach is realistic (see also the discussions in Section VI-B).

---

[6]The same approach can also be applied for imperfect $\mathbf{R}$.

As each of the elements of $\mathbf{U}$ has a maximum modulus of 1, $\bar{\mathbf{U}}$ can be chosen as the first $\alpha N_{RF}$ columns of $\mathbf{U}$. From these discussions, one can understand that the solution obtained in Section III is useful for designing the hybrid analog-digital channel estimator with $N_{RF} < N$.

**Optimality**: If $\text{rank}(\mathbf{W}^\star) \leq N_{RF}$ and $\alpha = 1$, then the optimal solution of (22) becomes $\bar{\mathbf{U}} = \mathbf{U}$ and $\bar{\mathbf{W}} = \mathbf{DV}^H$. This is due to the fact that the objective function of (22) is lower bounded by the minimum value of (5). However, as $\text{rank}(\mathbf{W}^\star) \leq N_{RF}$ may not hold always, choosing $\bar{\mathbf{U}}$ as the first $\alpha N_{RF}$ columns of $\mathbf{U}$ is not optimal for an arbitrary parameter settings.

For the given $\bar{\mathbf{U}}$, (22) and (23) can be solved exactly as in Section III. After solving (22), we will have the following normalized MSE

$$\bar{\xi} = \frac{1}{N}\text{tr}\{\bar{\boldsymbol{\xi}}\}. \tag{25}$$

Note that one can preselect $\bar{\mathbf{U}}$ different from the current paper. However, as the focus of the paper is to study the training-throughput tradeoff of the proposed hybrid analog-digital architecture, we utilize this $\bar{\mathbf{U}}$. And as will be detailed in the next subsection, the approach of examining the tradeoff between the training duration and its corresponding throughput can still be extended for other choices of $\bar{\mathbf{U}}$.

### B. Training-Throughput Tradeoff

In this section, we study the training-thoughput tradeoff for single user case when $N_{RF} < N$ and $\mathbf{R}$ is known perfectly. Data transmission can take place either in the downlink or uplink channel. As the BS has antenna arrays, it applies beamforming both in the downlink and uplink channels. By doing so, we will have the following estimated signals

$$\tilde{x} = \mathbf{h}^H\mathbf{b}x + n_d \quad \text{Downlink channel}, \quad \tilde{x} = \mathbf{a}^H(\mathbf{h}x + \mathbf{n}_d) \quad \text{Uplink channel} \tag{26}$$

where $x$ is the transmitted signal which is assumed to have $\text{E}|x|^2 = 1$, $\tilde{x}$ is the estimate of $x$, $\mathbf{b}$ ($\mathbf{a}^H$) is the precoding (decoding) vector, and $n_d$ and $\mathbf{n}_d$ are additive noises in the downlink and uplink channels, respectively with $n_d \sim \mathcal{CN}(0, \sigma_d^2)$ and $\mathbf{n}_d \sim \mathcal{CN}(\mathbf{0}, \sigma_d^2 \mathbf{I})$.

Note that the current paper examines both the training and data transmission phases. In some cases, it may be required to utilize different SNRs in these two phases. Specifically, the channel estimation phase may need to have high SNR to obtain better performance in the beamforming phase. One approach of controlling the SNR is by assigning different transmission powers in these two phases (i.e., the noise variance is fixed), and the other approach is to set the same transmission powers in both training and data transmission phases, and assume different noise variances. In the current paper, we have followed the latter approach just for mathematical convenience and better flow of the text. Nevertheless, in practice, the noise variance would be the same both in the channel estimation and beamforming phases. The desired SNR levels will, therefore, be maintained just by controlling the transmission power.

When $K = 1$, the downlink and uplink ZF approaches turn out to be maximum ratio transmission (MRT) and maximum ratio combining (MRC) approaches, respectively which can be designed as [23]

$$\mathbf{b} = \kappa \bar{\mathbf{h}} \quad \text{DL} - \text{MRT}, \quad \mathbf{a} = \bar{\mathbf{h}}^H \quad \text{UL} - \text{MRC} \tag{27}$$

where $\kappa = (\text{E}\{\bar{\mathbf{h}}^H \bar{\mathbf{h}}\})^{-1/2}$ is chosen to ensure that the average transmitted power in the uplink and downlink channels are the same (i.e., unity), and DL-MRT and UL-MRC denotes the downlink MRT and uplink MRC approaches, respectively.

According to [23], the transmitter and receiver vectors $\mathbf{b}$ and $\mathbf{a}$ can be implemented with a hybrid analog-digital architecture without performance loss when $N_{RF} \geq K$. For this reason, $N_{RF} = K = 1$ is sufficient for the transmission phase. Under this setting, the downlink and uplink average SNRs become

$$\text{E}\{\gamma\} = \frac{\text{E}|\mathbf{h}^H \mathbf{b} x|^2}{\text{E}|n_d|^2} = \frac{|t|^2 + \tilde{t}}{\sigma_d^2 \bar{t}} \quad \text{DL} - \text{MRT}, \quad \text{E}\{\gamma\} = \frac{\text{E}|\mathbf{a}^H \mathbf{h} x|^2}{\text{E}|\mathbf{a}^H \mathbf{n}_d|^2} = \frac{|t|^2 + \tilde{t}}{\sigma_d^2 \bar{t}} \quad \text{UL} - \text{MRC} \tag{28}$$

where

$$\text{E}\{\bar{\mathbf{h}}^H \bar{\mathbf{h}}\} = \bar{t}, \quad \text{E}\{|\mathbf{h}^H \bar{\mathbf{W}} \bar{\mathbf{n}}|^2\} = \tilde{t}, \quad \text{E}\{\mathbf{h}^H \bar{\mathbf{W}} \bar{\mathbf{U}}^H \mathbf{h}\} = t \tag{29}$$

with $t = \text{tr}\{\mathbf{R}\mathbf{M}\}$, $\tilde{t} = \sigma_t^2 \text{tr}\{\bar{\mathbf{W}}^H \mathbf{R} \bar{\mathbf{W}}\}$, $\bar{t} = \text{tr}\{\mathbf{M}\mathbf{R}\mathbf{M}^H\} + \sigma_t^2 \text{tr}\{\bar{\mathbf{W}}^H \bar{\mathbf{W}}\}$ and $\mathbf{M} = \bar{\mathbf{W}} \bar{\mathbf{U}}^H$. In

particular, when $\sigma_d^2 \ll 1$ and $\mathbf{h}$ is an i.i.d Rayleigh fading channel with $\mathbf{R} = \mathbf{I}$, we will have

$$\mathrm{E}\{\gamma\}_{\sigma_d^2 \ll 1} \approx \frac{\mathrm{tr}\{\bar{\mathbf{U}}^H \mathbf{R}^2 \bar{\mathbf{U}}(\bar{\mathbf{U}}^H \mathbf{R} \bar{\mathbf{U}})^{-1}\}}{\sigma_d^2} = \frac{\alpha}{\sigma_d^2}, \ \mathrm{DL-MRT \ and \ UL-MRC}. \tag{30}$$

With the above average SNRs, we get the following average normalized throughput[7]

$$Th = \mathrm{E}\left\{\frac{(T_c - \alpha T_s)}{T_c} \log_2(1 + \gamma)\right\} \leq \frac{(T_c - \alpha T_s)}{T_c} \log_2(1 + \mathrm{E}\{\gamma\}) \tag{31}$$

where the second inequality employs Jensen's inequality as $\log(1 + x)$ is a concave function [55][8].

As we can see from (28), $\mathrm{E}\{\gamma\}$ depends on $t$, $\tilde{t}$ and $\bar{t}$ which are functions of $\bar{\mathbf{U}}$. This validates that $\mathrm{E}\{\gamma\}$ depends on the training duration $\alpha$. In particular, increasing $\alpha$ increases $\mathrm{E}\{\gamma\}$ (easily seen from (30)) but it decreases $(T_c - \alpha T_s)$. This shows that the average throughput of the considered hybrid architecture is not necessarily the same for all training periods. Hence, $\alpha$ must be optimized to solve (1) as

$$\max_{\alpha} \frac{(T_c - \alpha T_s)}{T_c} \log_2(1 + \mathrm{E}\{\gamma\}). \tag{32}$$

To get better insight about the optimal $\alpha$, we consider the following lemma.

*Lemma 2*: When $N_{RF} = 1$, MRT (MRC) beamforming is employed and each of the elements of $\mathbf{h}$ is an i.i.d Rayleigh fading, (32) is strictly concave optimization problem where the optimal $\alpha$ is unique and can be obtained by simple bisection search method. Furthermore, for an arbitrary $\mathbf{h}$ with known covariance matrix $\mathbf{R}$ with $\left(\sum_{i=1}^{\alpha} \frac{g_i^2}{g_i + \sigma_t^2}\right) \gg \sigma_t^2$ and $\mathrm{E}\{\gamma\} \gg 1$ (i.e., at very high SNR), (32) is also a concave optimization problem where the optimal $\alpha$ can be found from

$$(\beta - \alpha)\log_2[\bar{f}(\alpha + 1)] - (\beta - \alpha + 1)\log_2[\bar{f}(\alpha)] = \log_2(\gamma_{max}) \tag{33}$$

where $g_i$ are the eigenvalues of $\mathbf{R}$ sorted in decreasing order (i.e., $g_1 \geq g_2 \geq, \cdots, \geq g_N$), $\gamma_{max} = \frac{g_1}{\sigma_d^2}$, $\bar{f}(\alpha) = \sum_{i=1}^{\alpha} \delta_i$ and $\delta_i = \frac{g_i}{g_1}$ with $0 \leq \delta_i \leq 1$.

*Proof:* See Appendix A ∎

It is evident that $\gamma_{max}$ increases as the spatial correlation of the UE's channel increases. From (33), we can thus notice that the optimal $\alpha$ decreases as the channel correlation increases. In other words, maximum training duration is required when the UE's channel coefficients are independent. This fact has also been validated in the numerical and simulation sections.

---

[7]Since $\mathrm{E}\{\gamma\}$ is the same in both the downlink and uplink channels, the average normalized throughput is the same in both of the channels.

[8]As will be demonstrated in the simulation section, this bound is tight.

As can be seen from Fig. 1, both channel estimation and data transmission operations are executed in each frame. In the $T_c - \tau$ duration, data transmission can be performed by three approaches. In the first approach, each of the $T_c - \tau$ duration is partitioned for both uplink and downlink channel data transmissions. In the second approach, either of the uplink or downlink data transmission takes place alternatively in each frame. The third approach could be a combination of the first and second approaches. In all these approaches, since $\mathrm{E}\{\gamma\}$ is the same in the downlink and uplink channels, the optimal $\alpha$ of (32) is the same for all of these three transmission approaches.

## V. CHANNEL ESTIMATION AND TRAINING-THROUGHPUT TRADEOFF FOR MULTIUSER WHEN $N_{RF} < N$

The key difference between the single user and multiuser cases is that the former does not experience interference while the latter does. This interference arises both during channel estimation and data transmission phases. In the following, first we provide our channel estimation approach for multiuser system employing hybrid analog-digital architecture, and then we utilize the estimated channel to study the tradeoff between training duration and throughput.

### A. Channel Estimation

Like in Section IV, the channel estimation phase can be performed both for perfect and imperfect $\mathbf{R}_k$ cases. To remove the effect of interference between the channel vectors of $K$ UEs, we assume that $\alpha$ is a multiple of $K$. And under this setting, the channel of each UE can be decoupled by allowing $K$ orthogonal pilots for fixed analog channel estimator part. In this regard, we suggest to fix $\bar{\mathbf{U}}_i$ (i.e., the analog channel estimator matrix) constant for the $i$th set of $K$ trainings. By doing so, one can get

$$\tilde{\mathbf{h}}_{ki} = \bar{\mathbf{U}}_i^H (\mathbf{h}_k + \tilde{\mathbf{n}}_k) \tag{34}$$

where $\tilde{\mathbf{n}}_k$ is as defined in (3). Now if we apply $\alpha = \theta K$ trainings and introduce the digital channel estimator of the $k$th UE $\bar{\mathbf{W}}_k$, we can express the estimate of $\mathbf{h}_k$ as

$$\bar{\mathbf{h}}_k = \bar{\mathbf{W}}_k^H \tilde{\mathbf{h}}_k \tag{35}$$

where $\tilde{\mathbf{h}}_k = [\tilde{\mathbf{h}}_{k1}^T, \tilde{\mathbf{h}}_{k2}^T, \cdots, \tilde{\mathbf{h}}_{k\theta}^T]^T$ and $\bar{\mathbf{U}} = [\bar{\mathbf{U}}_1, \bar{\mathbf{U}}_2, \cdots, \bar{\mathbf{U}}_\theta]$. The design of $\bar{\mathbf{U}}$ and $\bar{\mathbf{W}}_k$ can be examined like in (22) and (23) for perfect and imperfect $\mathbf{R}_k$ cases, respectively. Furthermore, it can be shown that the joint optimization of $\bar{\mathbf{U}}$ and $\bar{\mathbf{W}}_k$ will result to a non-convex problem. For this reason, we preselect $\bar{\mathbf{U}}$ and optimize $\bar{\mathbf{W}}_k$ for fixed $\bar{\mathbf{U}}$ like in the above section. To this end, we choose $\bar{\mathbf{U}}$ as the first $\theta N_{RF}$ columns of $\mathbf{U}$, where

$$\mathbf{W}^\star = [\mathbf{W}_1^\star, \mathbf{W}_2^\star, \cdots, \mathbf{W}_K^\star] = \mathbf{U}\mathbf{D}\mathbf{V}^H, \tag{36}$$

$\mathbf{D}$ is a diagonal matrix containing the non-zero singular values of $\mathbf{W}^\star$ arranged in decreasing order, and $\mathbf{W}_k^\star$ are the solutions obtained from (6) for perfect $\mathbf{R}_k$ case (by solving (12) for imperfect $\mathbf{R}_k$ case).

Once the channel estimation is performed, we will have the following normalized total MSE

$$\bar{\xi} = \frac{1}{KN} \sum_{k=1}^{K} \mathrm{tr}\{\bar{\boldsymbol{\xi}}_k\}. \tag{37}$$

### B. Training-Throughput Tradeoff

This subsection discusses the training-throughput tradeoff for multiuser scenario with perfect $\mathbf{R}_k$. For better exposition of the proceeding discussions, here we also assume that $\alpha$ is a multiple of $K$. Under this assumption and after doing some mathematical manipulations, we can rewrite the combined estimated channel of all UEs $\bar{\mathbf{H}} \triangleq [\bar{\mathbf{h}}_1, \bar{\mathbf{h}}_2, \cdots, \bar{\mathbf{h}}_K]$ obtained from (35) as

$$\bar{\mathbf{H}} = \mathbf{M}\mathbf{H}_d + \bar{\mathbf{W}}\bar{\mathbf{N}}_d \tag{38}$$

where $\mathbf{M} = [\mathbf{M}_1, \mathbf{M}_2, \cdots, \mathbf{M}_K]$ with $\mathbf{M}_k = \bar{\mathbf{W}}_k^H \bar{\mathbf{U}}^H$, $\bar{\mathbf{W}} = [\bar{\mathbf{W}}_1, \bar{\mathbf{W}}_2, \cdots, \bar{\mathbf{W}}_K]$, $\mathbf{H}_d = \mathrm{blkdiag}(\mathbf{h}_1, \mathbf{h}_2, \cdots, \mathbf{h}_K)$ and $\bar{\mathbf{N}}_d = \mathrm{blkdiag}(\bar{\mathbf{n}}_1, \bar{\mathbf{n}}_2, \cdots, \bar{\mathbf{n}}_K)$. If we again assume that the channels of all UEs are uncorrelated which is usually the case in practice, we will have

$$\mathrm{E}\{\bar{\mathbf{H}}^H \bar{\mathbf{H}}\} = \mathrm{diag}(\bar{t}_1, \bar{t}_2, \cdots, \bar{t}_K), \quad \mathrm{E}\{(\mathbf{M}\mathbf{H}_d)^H \mathbf{H}\} = \mathrm{E}\{\mathbf{H}^H \mathbf{M}\mathbf{H}_d\} = \mathrm{diag}(t_1, t_2, \cdots, t_K) \tag{39}$$

where $\bar{t}_k = \mathrm{tr}\{\mathbf{R}_k \mathbf{M}_k^H \mathbf{M}_k\} + \sigma_t^2 \mathrm{tr}\{\bar{\mathbf{W}}_k^H \bar{\mathbf{W}}_k\}$ and $t_k = \mathrm{tr}\{\mathbf{R}_k \bar{\mathbf{W}}_k^H \bar{\mathbf{U}}^H\}$.

The data transmission phase can be either in the uplink or dowlink channel. To this end, we employ the well known ZF approach in both the downlink and uplink transmission phases, respectively as

$$\mathbf{B} = \tilde{\kappa} \bar{\mathbf{H}}(\bar{\mathbf{H}}^H \bar{\mathbf{H}})^{-1} \ \ \mathrm{DL-ZF}, \quad \mathbf{A} = (\bar{\mathbf{H}}^H \bar{\mathbf{H}})^{-1} \bar{\mathbf{H}}^H \ \ \mathrm{UL-ZF} \tag{40}$$

where $\tilde{\kappa} = (\mathrm{E}\{\mathrm{tr}\{(\widehat{\mathbf{H}}^H \widehat{\mathbf{H}})^{-1}\}\})^{-1/2}$ is introduced to maintain the BS average transmitted power to unity, and DL-ZF (UL-ZF) denotes the downlink (uplink) zero forcing.

With these beamfoming matrices and using (39), the downlink and uplink average SNR of the $k$th UE are given as

$$\mathrm{E}\{\gamma_k\} = \frac{\tilde{\kappa}^2}{\sigma_d^2}\left[\mathbf{H}^H \bar{\mathbf{H}}(\bar{\mathbf{H}}^H \bar{\mathbf{H}})^{-2}\bar{\mathbf{H}}^H \mathbf{H}\right]_{k,k} = \frac{|t_k|^2}{\sigma_d^2 |\bar{t}_k|^2 (\sum_{i=1}^K \bar{t}_i^{-1})}, \quad \mathrm{DL-ZF \ and \ UL-ZF}. \tag{41}$$

And the average normalized throughput becomes

$$Th = \mathrm{E}\left\{\frac{(T_c - \alpha T_s)}{T_c}\sum_{i=1}^K \log_2(1+\gamma_i)\right\} \le \frac{(T_c - \alpha T_s)}{T_c}\sum_{i=1}^K \log_2(1+\mathrm{E}\{\gamma_i\}). \tag{42}$$

Like in the single user case, $\mathrm{E}\{\gamma_k\}$ depends on $t_k$ and $\bar{t}_k$ which are functions of $\bar{\mathbf{U}}$. This validates that $\mathrm{E}\{\gamma_k\}$ depends on the training duration $\alpha$ which may need to be optimized for maximizing $Th$ as

$$\max_\alpha \frac{(T_c - \alpha T_s)}{T_c}\sum_{i=1}^K \log_2(1+\mathrm{E}\{\gamma_i\}). \tag{43}$$

For better exposition on the optimal $\alpha$ for multiuser systems, we consider the following lemma.

*Lemma 3*: When $\alpha = \theta K$ and the channel of each UE is i.i.d zero mean Rayleigh fading (i.e., $\mathbf{R}_k = q_k\mathbf{I}$), (43) is a strictly concave optimization problem where the optimal $\theta$ is unique and can be obtained by simple bisection search method. Specifically, when $\mathrm{E}\{\gamma_k\} \gg 1$ (i.e., at high SNR), the optimal $\theta$ satisfies

$$\frac{\beta}{\theta} - K\log_2(\theta) = K\left[1 + \log_2\left(\frac{\bar{q}}{\sigma_d^2}\right)\right] \tag{44}$$
$$\Rightarrow K\left[1 + \log_2(\theta) + \log_2\left(\frac{\bar{q}}{\sigma_d^2}\right)\right] = \frac{\beta}{\theta}$$

where $\bar{q} = \left(\sum_{i=1}^K \frac{q_i + \sigma_t^2}{q_i^2}\right)^{-1}$.

*Proof:* See Appendix B ∎

From (44) one can notice that the optimal $\theta$ decreases when $\bar{q}$ increases (i.e., the optimal training duration increases as the SNR decreases). This fact has also been demonstrated both in the numerical and simulation results by considering practically relevant SNR regions.

For the fixed $\bar{q}$, $\sigma_d^2$ and $\theta$ (i.e., independent of $K$), one can notice from (44) that $\beta$ may need to increase as $K$ increases to maintain the above equality which is required at optimality. From this discussion, we can understand that the maximum possible value of $\beta$ for the mmWave and microwave frequency bands may vary from one scenario to another (i.e., cannot be determined a priori). However, the minimum value of $\beta$ should be the same as that of the optimal training duration obtained from the fully digital architecture case. This is because the digital architecture has more degrees of freedom (which consequently help utilize the minimum training duration) than that of the hybrid architecture.

In the multiuser setup, providing analytical proof showing that (43) is concave for general settings is not trivial as the $\mathrm{E}\{\gamma_k\}$ of (28) depends on $K$, $\mathbf{R}_k$, $\sigma_t^2$ and $\sigma_d^2$. Nevertheless, since computing $\mathrm{E}\{\gamma_k\}$ numerically is quite simple for the given $\mathbf{R}_k$, it is still possible to get the optimal $\theta$ by examining $Th$ numerically. And as will be demonstrated in the numerical and simulation sections, problem (43) is still concave for practically relevant parameter settings.

Note that the analog channel estimation matrix $\bar{\mathbf{U}}_i$ is realized using RF electronic components which usually introduce some delays when the corresponding matrix is switched from $\bar{\mathbf{U}}_i$ to $\bar{\mathbf{U}}_{i+1}$. For instance, in the single user setup, $\bar{\mathbf{U}}_i$ may need to be updated every $T_s$. However, one can still treat $\tilde{\mathbf{y}}_i$ of (17) as the received signal obtained at each measurement period $T_m$ which consists of $T_s$ and possible delay $T_d$ (i.e., $T_m = T_s + T_d$). And, from the discussions of Sections IV and V, one can learn that the optimal training duration does not depend on the specific values of $T_m(T_s)$. For this reason, the introduction of delays in the analog components will not affect the analysis of the current paper. We would also like to point out that adaptive compressive sensing method has been utilized in [24] where the authors employ the number of measurements (observations) to facilitate the channel estimation. This paper also follows the assumption that the analog beamforming matrix be varied for each measurement (see (9) of [24]).

**Extension for the case where** $\alpha \neq \theta K$: The channel estimation and training-throughput tradeoff discussions assume that $\alpha$ is a multiple of $K$. However, it may also be interesting to consider the case where $\alpha \neq \theta K$. In such a case, we propose to allow only the first $\bar{K} = \text{rem}(\frac{\alpha}{K})$ UEs transmit their orthogonal pilots during the channel estimation phase when $(\theta - 1)K < \alpha < \theta K, \forall \theta$. And these $\bar{K}$ UEs can be selected based on their equivalent channel gains (i.e., $\text{tr}\{\bar{\mathbf{U}}_\theta^H \mathbf{R}_k \bar{\mathbf{U}}_\theta\}$). By doing so, each UE's channel vector will not experience interference from the other UE's channel. To compute the optimal $\alpha$, we suggest to use a two step approach, i.e., first we compute $\theta$ using (43) (fast searching), then we tune $\alpha$ in the range $(\theta - 1)K \leq \alpha \leq (\theta + 1)K$ (tuning). By doing so, the complexity of searching the optimal $\alpha$ can be reduced significantly[9].

**Complexity**: The computational complexity of the considered training-throughput tradeoff arises from solving the channel estimator discussed in Section III-C and the complexity of Appendix A (B) (i.e., computation of the optimal training time). However, since the optimal solution computed from Appendix A (B) can be obtained using bisection search which has negligible computational load [49], the overall complexity of the training-throughput tradeoff study is the same as the one discussed in Section III-C.

We would like to recall here that the training-throughput tradeoff is studied by utilizing the normalized throughputs $Th$ given in (31) and (42). Since each of these expressions has an outer expectation term, optimizing $Th$ as it is appears to be challenging. For this reason, we employ Jensen's inequality to get the upper bound of the normalized throughput which has convenient mathematical structure for studying the training-throughput tradeoff. Furthermore, we have demonstrated by computer simulation that the examined throughput is tight (see Section VI-A). However, this simulation result cannot act as a general proof to ensure the tightness of the upper bound throughput. Therefore, studying the training-throughput tradeoff by considering the exact (theoretically tight bound) throughput is left for future research.

Up to now, we assume that the covariance matrix of each UE is known either perfectly or imperfectly.

---

[9]As will be clear in the numerical and simulation sections, the optimal $\alpha$ decreases as the rank of $\mathbf{R}_k$ decreases. This fact again can be exploited to get the optimal $\alpha$ with reduced complexity.

In fact, when $\mathbf{R}_k$ are known and available a priori (for example in the i.i.d Rayleigh fading channel environment), there is no need to estimate the covariance matrix. However, this may not happen for all scenarios such as an indoor environment. This validates the need to estimate the covariance matrix. In this regard, we have provided a brief summary of the sample channel covariance matrix estimator for each UE when $N_{RF} < N$ in Appendix C[10].

**Extension to Multiuser MIMO Case:** In the analysis of this paper, it is assumed that each of the UEs is equipped with single antenna. However, the deployment of multiple antennas are adopted in different existing standards. In microwave frequency bands, as each of the UEs is likely equipped with few number of antennas, the conventional digital architecture is still a reasonable solution. For such a case, one can apply the training-throughput tradeoff study of the current paper just by treating the number of antennas as the total number of "virtual single antenna UEs". However, at mmWave frequencies, a hybrid architecture can be deployed at each UE. For such a case, one can study the training-throughput tradeoff by considering two systems: The first system assumes that each UE utilizes prefixed analog channel estimation and beamforming matrices. This analog matrix can be selected by exploiting the history of channel covariance statistics (such an approach drastically reduces the power consumption and computational load at each UE). For this system, one can notice that the analysis of the current paper can be applied straightforwardly. The second system does not impose any a priori assumption on the analog architecture of each UE. In general, such an assumption leads to a superior performance at the expense of increased complexity. And studying the training-throughput tradeoff for this system is still an open problem and is left for future work.

## VI. NUMERICAL AND MONTE CARLO SIMULATION RESULTS

This section provides numerical and Monte Carlo simulation results of the normalized MSE and throughput expressions. In this regard, we set $N_{RF} = K$, $\sigma_t^2 = \sigma_d^2$, $\beta = 2KT_o$, $\alpha = \theta K$ and we vary $T_o$, $N$, $K$, $\theta$, $\mathbf{R}_k$, where $T_o$ is introduced for convenience. To this end, $\mathbf{R}_k$ are taken from a widely used exponential correlation model as $\mathbf{R}_k = \rho_k^{|i-j|}$, where $0 \leq \rho_k < 1$. We have used exponential correlation model because of the following two reasons[11]. First, the exponential model is physically reasonable in a way that the correlation between two transmit antennas decreases as the distance between them increases [44]. Second, this model is a widely used antenna correlation model for an urban area communications where traffic is usually congested [56]. We utilize the SNR which is defined as $\frac{\mathrm{E}\{|x_k|^2\}}{\sigma_d^2}$ (i.e., the multiuser version of $x$ in (26)) and normalized training duration (i.e., $\theta$).

---

[10]We would like to point out here that as the analysis of the current paper depends on $\mathbf{R}_k$ which is the same for all sub-carriers, the training-throughput tradeoff study is also valid for frequency selective channels.

### A. Numerical Results

This subsection presents the numerical results obtained from the normalized MSE and throughput expressions for the scenario where $\mathbf{R}_k$ are known perfectly.

*1) Single User:* The achieved normalized MSE $\bar{\xi}$ for $N = 64$ and $T_o = 64$, and different $\theta$, $\rho = \rho_1$ and SNR is plotted in Fig. 2. As can be seen from this figure, increasing $\theta$ improves the quality of the channel estimator by reducing the MSE which is expected. On the other hand, for a given operating SNR, the required training duration to achieve a given target MSE decreases as $\rho$ increases. This result is expected because as $\rho$ increases, $\frac{\lambda_i}{\lambda_1}$ with $\lambda_1 \geq \lambda_2, \cdots \lambda_N$ decreases with $i$ which consequently help to reduce the training duration. Fig. 3 shows the achievable throughput versus training duration for different SNR and $\rho$.

As can be seen from this figure, despite the conventional channel estimation approach, the optimal training duration of the proposed hybrid analog-digital channel estimator is not necessarily the same for all $\rho$. In particular, the optimal training duration decreases as the rank of the channel covariance matrix decreases (i.e., as $\rho$ increases). As an example, when SNR=10dB, the optimal training durations are around $4T_s, 12T_s$ and $20T_s$ for $\rho = 0.9, 0.5$ and $0$ (i.e., i.i.d Rayleigh fading channel), respectively. On the other hand, less training period is employed when the UE's channel is highly correlated which consequently results a higher normalized throughput. These results confirm that the training period should be selected carefully when the BS has less number of RF chains than that of antennas.

---

[11]We have also applied the analysis of the current paper for ULA and uniform planar array (UPA) channel models, and have found similar training-throughput tradeoff characteristics to that of the current paper. More detailed results can be found in the next section.

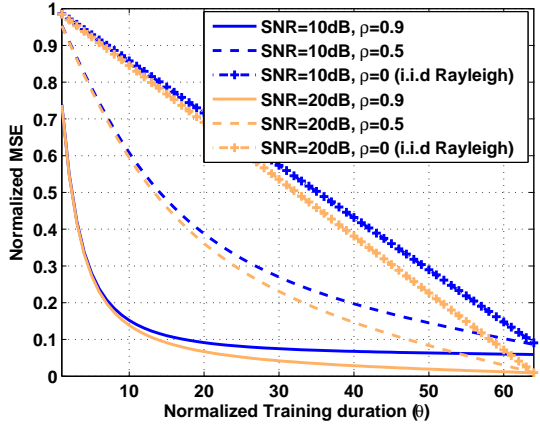

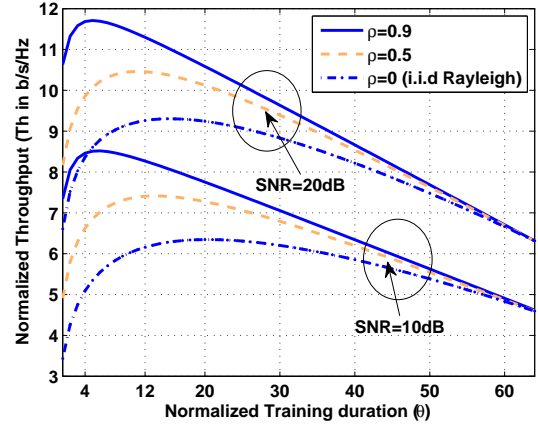

Fig. 2. Normalized MSE versus training duration of the hybrid analog-digital channel estimator for single user case.

Fig. 3. Normalized throughput of the hybrid analog-digital channel estimation and beamforming for single user case.

*2) Multiuser:* This subsection examines the training-throughput tradeoff for multiuser case. To this end, we set $K = 4$ $T_o = 32$, normalized SNR of all users as $\{10, 20\}$dB, and three sets of $\boldsymbol{\rho}$ as $\boldsymbol{\rho}_1 = [0, 0, 0, 0]$, $\boldsymbol{\rho}_2 = [0.5, 0.5, 0.3, 0.5]$ and $\boldsymbol{\rho}_3 = [0.85, 0.9, 0.75, 0.95]$. Fig. 4 shows the througput versus normalized training for these settings. As we can see from this figure, the optimal training duration depends on the channel coherence time. Furthermore, like in the single user case, the optimal $\theta$ for multiuser case decreases as the correlation coefficients of all UEs increase. On the other hand, for a given $\boldsymbol{\rho}$, the optimal training duration is not necessarily the same for all SNR regions. For instance, when $\boldsymbol{\rho} = \boldsymbol{\rho}_2$, the optimal $\theta$s are around $18$ and $25$ for the $20$dB and $10$dB normalized SNR values, respectively. This observation validates the optimal structure of $\theta$ which is provided in *Lemma 3* for $\boldsymbol{\rho} = \mathbf{0}$.

### B. Simulation results

This subsection demonstrates the theoretical normalized MSE and throughput expressions via Monte Carlo simulations. All of the results are obtained by averaging 20000 channel realizations. For this subsection, we employ $N = 64$, $K = 4$ and $T_c = 256T_s$, and the training symbols $\mathbf{P}$ are taken from the discrete Fourier transform (DFT) matrix with appropriate size.

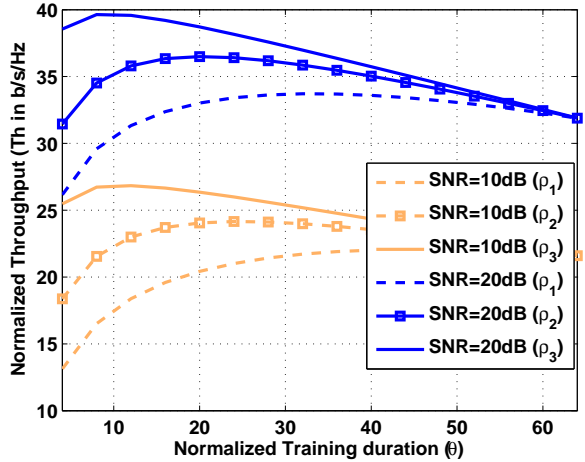

Fig. 4. Normalized throughput of the hybrid analog-digital channel estimation and beamforming for multiuser case.

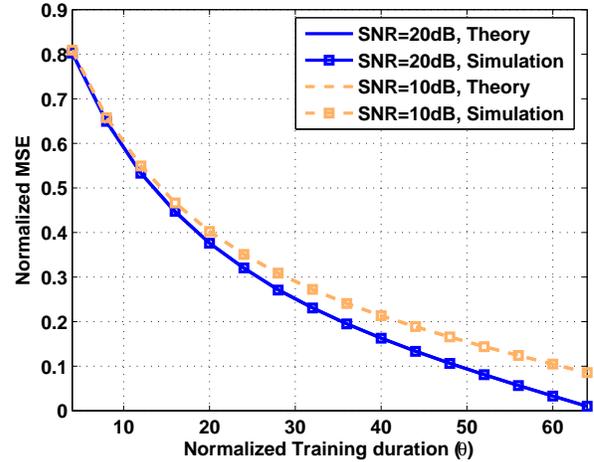

Fig. 5. Theoretical and simulated normalized MSE of the hybrid analog-digital channel estimator with perfect $\boldsymbol{\rho}$.

*1) Channel Estimation:* The channel estimation phase is examined both for perfect and imperfect $\mathbf{R}_k$ scenarios. To this end, this subsection verifies the theoretical normalized MSE expressions, and validates the effectiveness of the proposed robust channel estimator compared to that of the non-robust one. Fig. 5 shows the theoretical and simulated normalized MSEs for different $\theta$ when the SNR is set to 10dB and 20dB, and $\mathbf{R}_k$ is known perfectly and is computed from $\boldsymbol{\rho} = [0.4, 0.6, 0.2, 0.7]$. As can be seen from this figure, the simulated normalized MSEs match that of the theoretical ones for both SNRs.

In the next simulation, we examine the effect of covariance matrix estimation error on the MSE of the robust and non-robust designs. Towards this end, we compare the MSE achieved by the robust and non-robust channel estimators. Fig. 6 shows the achieved normalized MSEs for the estimated values $\hat{\boldsymbol{\rho}}_1 = [0.40, 0.60, 0.20, 0.70]$, $\hat{\boldsymbol{\rho}}_2 = [0.75, 0.90, 0.70, 0.95]$ and $\hat{\boldsymbol{\rho}}_3 = [0.95, 0.96, 0.90, 0.98]$ while setting $\epsilon_k = \sqrt{N}$. And the true $\mathbf{R}_k^{1/2} = \hat{\mathbf{R}}_k^{1/2} + \boldsymbol{\Delta}_k$, where $\|\boldsymbol{\Delta}_k\| \leq \epsilon_k$. As can be seen from this figure, the robust design achieves less normalized MSE compared to that of the non-robust one, and its improvement is higher when $\boldsymbol{\rho}$ is higher. This is because as $\boldsymbol{\rho}$ is higher, the preselected $\bar{\mathbf{U}}$ in (22) will be closer to its optimal value which consequently helps to improve the performance of the robust channel estimator[12]. As expected, the lowest normalized MSE is achieved when $\boldsymbol{\rho}$ is known perfectly.

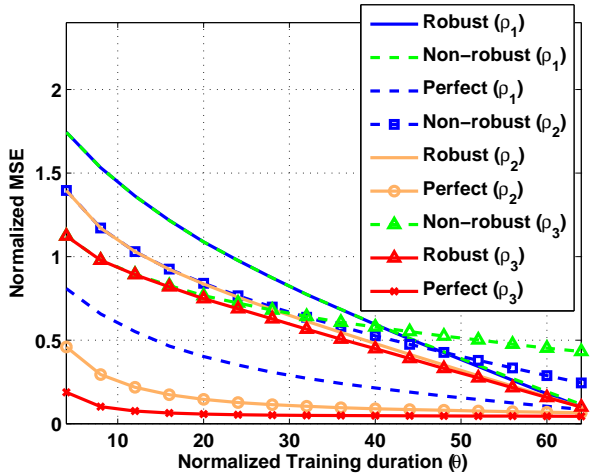

Fig. 6. Comparison of perfect, robust and non-robust hybrid analog-digital channel estimators.

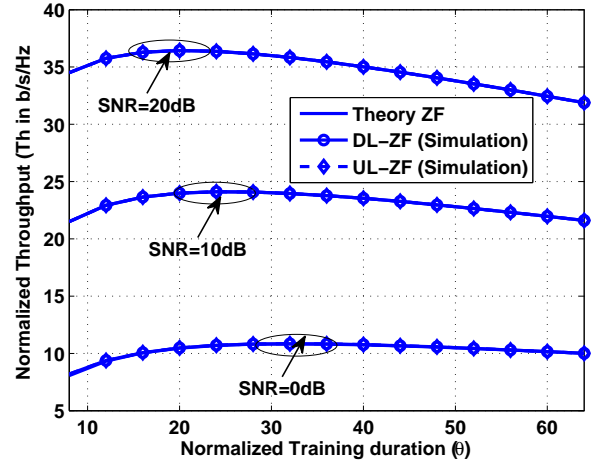

Fig. 7. Theoretical and simulated normalized throughput of hybrid analog-digital channel estimator and beamformer.

*2) Data Transmission:* In this subsection, the throughput expression given in (42) is verified via computer simulations. We also demonstrate the optimal training duration and examine the tightness of the average upper bound throughput given in (42) with simulations. In this regard, we choose $\boldsymbol{\rho} = [0.40, 0.60, 0.20, 0.70]$ and consider the case where $\mathbf{R}_k$ is known perfectly. To plot the simulation normalized throughput, first we compute the SNR from Monte Carlo simulation for each channel realizations, then we employ the first equality of (42) for the evaluation of the average normalized throughput.

Fig. 7 shows the theoretical and simulated normalized throughput versus training when $\mathbf{R}_k$ is known perfectly. This figure demonstrates that the bound derived in (42) is tight since the simulated normalized throughput is almost the same as that of the upper bound throughput for both the downlink and uplink ZF beamforming schemes[13]. On the other hand, the optimal $\theta$ that maximizes the average throughput decreases as the SNR increases which is expected. From Fig. 7, one can also realize that utilizing the solution of [32] (i.e., $\theta = 1$) yields a significant reduction in throughput.

---

[12]The non-robust channel estimator is the design that does not take into account $\epsilon_k$ in the design. And the perfect channel estimator corresponds to the case where $\epsilon_k = 0$.

[13]The cause of this throughput tightness is likely due to the fact that massive MIMO systems become less sensitive to the actual entries of the channel matrix which yields $\mathrm{E}\{\gamma\} \approx \gamma$ (i.e., channel hardening [4]).

Note that the simulation results of this paper assumes that $\bar{\mathbf{U}}$ is realized with sufficiently high (theoretically infinite) resolution phase shifters. When the hybrid architecture employs constant phase phase shifters and switches, each of the elements of $\bar{\mathbf{U}}$ can be realized with some accuracy only [23]. Specifically, as the required accuracy increases, the number of fixed phase phase shifters and switches increase. When the accuracy is set to $10^{-1}$ (i.e., the absolute difference between $\bar{\mathbf{U}}_{ij}$ and its approximated value), we have found almost the same MSE and throughput as those of the plots of this section. The simulation results demonstrating this fact has been omitted for conciseness.

## VII. Extra Simulation results for ULA Channels

Millimeter wave channels are expected to be specular and have low rank, and may not necessarily follow the same covariance structure as that of the exponential channel correlation models discussed in the above section. In this regard, this section provides some simulation results on the evaluation of the training throughput tradeoff for mmWave channels. To this end, we consider a ULA having different number of scatterers ($L_k$) (all the other settings are the same as that of the above section). The AOD of each UE is assumed to have arbitrary phases taken from the uniform distribution in $[0, \ 2\pi]$.

Fig. 8 shows the training versus throughput curves for different number of scatterers. All the other parameter settings are as in the first paragraph of Section VII of the revised manuscript. As can be seen from this figure, the optimal training duration increases as $L_k$ increases which is expected. This is because, the rank of the channel increases as $L_k$ increases which fits with the theoretical result provided in the above section.

## VIII. Conclusions

This paper considers hybrid analog-digital channel estimation and beamforming for multiuser massive MIMO systems with limited number of RF chains. Under these settings, first, we design novel TDD MMSE based hybrid analog-digital channel estimator by considering both perfect and imperfect channel covariance matrix cases. Then, we utilize the estimated channels to exploit beamforming for data transmission. Under the assumption of perfect channel covariance matrix, we show that there is a tradeoff between the training duration and achievable throughput when the number of RF chains is limited and hybrid analog-digital channel estimation and beamforming is applied. Specifically, we exploit the fact that the optimal training duration that maximizes the overall network throughput depends on the available number of RF chains, channel coherence time and covariance matrices of all UEs. We also show that the training time optimization problem can be formulated as a concave maximization problem for some system parameter settings where its global optimal solution can be obtained efficiently using existing tools. The analytical expressions are validated by performing extensive numerical and Monte Carlo simulations. The robustness of the proposed robust channel estimator is demonstrated using computer simulations.

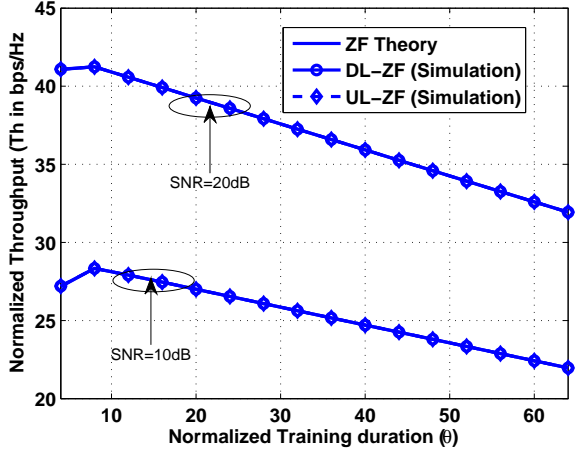

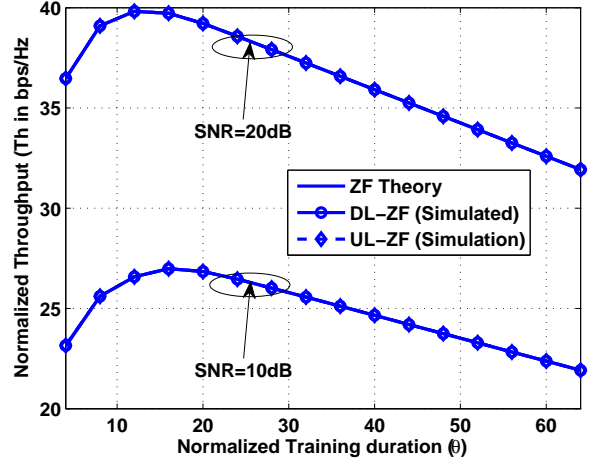

(a)

(b)

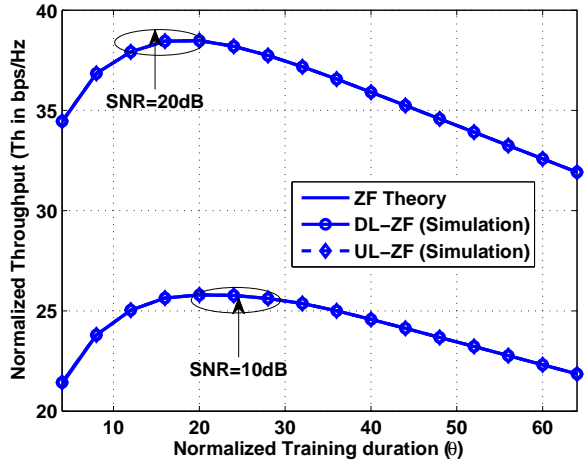

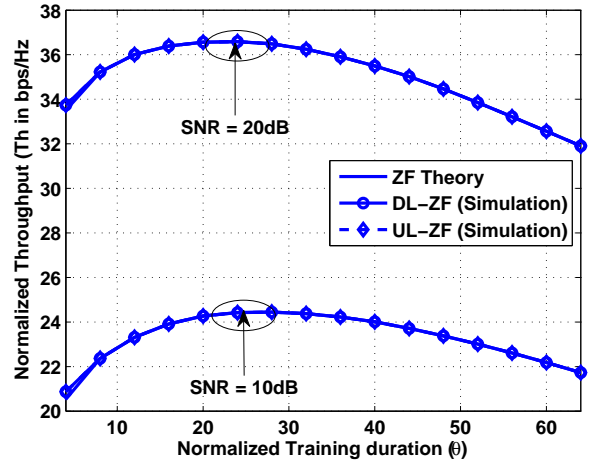

(c)

(d)

Fig. 8. Theoretical and simulated normalized throughput versus training of the hybrid analog-digital channel estimator and beamformer with ULA channel model having different number of scatterers ($L_k$). (a): $L_k = 4$, (b): $L_k = 12$, (c): $L_k = 24$, (d): $L_k = 64$.

## APPENDIX A

## PROOF OF *Lemma 2*

As can be seen from (28), $\mathrm{E}\{\gamma\}$ is the same for both MRC and MRT approaches. For this reason, this appendix provides the proof of *Lemma 1* when the BS has 1 RF chain and employ downlink data transmission. After some mathematical manipulations, the average SNR can be expressed as

$$\mathrm{E}\{\gamma\} = \frac{\mathrm{E}|\mathbf{h}^H \mathbf{b}|^2}{\sigma_d^2} = \frac{1}{\sigma_d^2} \frac{\left[ \left( \sum_{i=1}^{\alpha} \frac{g_i^2}{g_i + \sigma_t^2} \right)^2 + \sigma_t^2 \sum_{i=1}^{\alpha} \frac{g_i^2}{(g_i + \sigma_t^2)^2} \right]}{\sum_{i=1}^{\alpha} \frac{g_i^3 + \sigma_t^2 g_i^2}{(g_i + \sigma_t^2)^2}} \tag{45}$$

where $g_i$ are the eigenvalues of $\mathbf{R}$ sorted in decreasing order (i.e., $g_1 \geq g_2 \geq, \cdots, \geq g_N$).

### A. When $\mathbf{h}$ is an i.i.d Rayleigh fading channel

Under an i.i.d Rayleigh fading channel, we will have $g_1 = g_2 = \cdots, = g_N \triangleq g$ and $\mathrm{E}\{\gamma\}$ becomes

$$\mathrm{E}\{\gamma\} = \frac{1}{\sigma_d^2} \left( \frac{\alpha g^2 + \sigma_t^2}{g + \sigma_t^2} \right).$$

Thus, we can rewrite (32) as

$$\max_\alpha \frac{(\beta - \alpha)}{\beta} \log_2(\alpha a + b) \triangleq f(\alpha) \tag{46}$$

where $a = \frac{g^2}{\sigma_d^2 (g + \sigma_t^2)}$ and $b = \frac{\sigma_t^2}{\sigma_d^2 (g + \sigma_t^2)} + 1$. From the second order derivative of $f(\alpha)$ we get

$$\frac{d^2 f}{d\alpha^2} = -\frac{a}{\alpha a + b} \left( \frac{2}{\beta} + \frac{\beta - \alpha}{\beta} \frac{a}{\alpha a + b} \right). \tag{47}$$

As $a > 0$ and $b > 0$, $\frac{d^2 f}{d\alpha^2} < 0$ for $0 < \alpha \leq \beta$. Therefore, (46) is strictly concave function, and the optimal $\alpha$ is unique and obtained by applying simple bisection search approach [49], [57].

## B. Arbitrary $\mathbf{h}$

In the following, we show that (46) is strictly concave for an arbitrary $\mathbf{R}$ (i.e., $\mathbf{h}$) when $\left( \sum_{i=1}^{\alpha} \frac{g_i^2}{g_i + \sigma_t^2} \right) \gg$ $\sigma_t^2$ (i.e., at very high SNR). For such a setting, $\mathrm{E}\{\gamma\}$ of (45) can be expressed as

$$\mathrm{E}\{\gamma\} = \frac{1}{\sigma_d^2} \sum_{i=1}^{\alpha} g_i = \gamma_{max} \sum_{i=1}^{\alpha} \delta_i \tag{48}$$

where $\gamma_{max} = \frac{g_1}{\sigma_d^2}$, $\delta_i = \frac{g_i}{g_1}$ with $0 \leq \delta_i \leq 1$. We again assume that $\mathrm{E}\{\gamma\} \gg 1$. Under such assumption, we can reexpress the optimization problem (32) as

$$\max_{\alpha} \ (\beta - \alpha)[\log_2(\gamma_{max}) + \log_2(\bar{f}(\alpha))] \triangleq f(\alpha) \tag{49}$$

where $\bar{f}(\alpha) = \sum_{i=1}^{\alpha} \delta_i$. The first and second derivatives of $f(\alpha)$ can be obtained by applying a finite difference method. Upon doing so, we will have

$$\frac{df(\alpha)}{d\alpha} = (\beta - \alpha)[\log_2(\bar{f}(\alpha+1)) - \log_2(\bar{f}(\alpha))] - [\log_2(\gamma_{max}) + \log_2(\bar{f}(\alpha))] \tag{50}$$

$$\frac{d^2 f(\alpha)}{d\alpha^2} = (\beta - \alpha) \log_2[\bar{f}(\alpha+2)] - 2(\beta - \alpha + 1) \log_2[\bar{f}(\alpha+1)] + (\beta - \alpha + 2) \log_2[\bar{f}(\alpha)]$$

$$= (\beta - \alpha)(\log_2[\bar{f}(\alpha+2)] - \log_2[\bar{f}(\alpha+1)]) - (\beta - \alpha + 2)(\log_2[\bar{f}(\alpha+1)] - \log_2[\bar{f}(\alpha)])$$

$$\leq (\beta - \alpha)(\log_2[\bar{f}(\alpha+1)] - \log_2[\bar{f}(\alpha)]) - (\beta - \alpha + 2)(\log_2[\bar{f}(\alpha+1)] - \log_2[\bar{f}(\alpha)])$$

$$= -2(\log_2[\bar{f}(\alpha+1)] - \log_2[\bar{f}(\alpha)]) \leq 0 \tag{51}$$

where the fourth and last inequalities are because of the fact that $\delta_i$ are arranged in decreasing order and $\bar{f}(\alpha) \leq \bar{f}(\alpha+1) \leq \bar{f}(\alpha+2)$. As $\frac{d^2 f(\alpha)}{d\alpha^2} \leq 0$, $f(\alpha)$ is again concave function and the optimal $\alpha$ can be obtained by equating $\frac{df(\alpha)}{d\alpha} = 0$ as in (33).

## Appendix B

## Proof of *Lemma 3*

From (39), one can obtain

$$\bar{t}_k = \mathrm{tr}\{\mathbf{R}_k \bar{\mathbf{W}}_k^H \bar{\mathbf{U}}^H\} = \mathrm{tr}\{\bar{\mathbf{U}}^H \mathbf{R}_k^2 \bar{\mathbf{U}}(\bar{\mathbf{U}}^H \mathbf{R}_k \bar{\mathbf{U}} + \sigma_t^2 \mathbf{I})\}$$

$$t_k = \mathrm{tr}\{\mathbf{R}_k \bar{\mathbf{W}}_k^H \bar{\mathbf{U}}^H\} + \sigma_t^2 \mathrm{tr}\{\mathbf{W}_k^H \mathbf{R}_k \mathbf{W}_k\}$$

$$= \mathrm{tr}\{\bar{\mathbf{U}}^H \mathbf{R}_k \bar{\mathbf{U}}(\bar{\mathbf{U}}^H \mathbf{R}_k \bar{\mathbf{U}} + \sigma_t^2 \mathbf{I})^{-1} \bar{\mathbf{U}}^H \mathbf{R}_k^2 \bar{\mathbf{U}}(\bar{\mathbf{U}}^H \mathbf{R}_k \bar{\mathbf{U}} + \sigma_t^2 \mathbf{I})^{-1}\} + \sigma_t^2 \mathrm{tr}\{\bar{\mathbf{U}}^H \mathbf{R}_k^2 \bar{\mathbf{U}}^H (\bar{\mathbf{U}}^H \mathbf{R}_k \bar{\mathbf{U}} + \sigma_t^2 \mathbf{I})^{-2}\}.$$

When $\mathbf{h}_k$ are i.i.d zero mean Rayleigh fading, $\mathbf{R}_k$ can be represented by a scaled identity matrix (i.e., $\mathbf{R}_k = q_k \mathbf{I}$). Consequently, $t_k$ and $\bar{t}_k$ can be expressed as $t_k = \bar{t}_k = \theta \frac{q_k^2}{q_k + \sigma_t^2}$. By plugging these expressions into (39), one can represent the average SNR of the $k$th UE as

$$\mathrm{E}\{\gamma_k\} = \frac{\bar{q}}{\sigma_d^2}\theta. \tag{52}$$

where $\bar{q}$ is as defined in *Lemma 3*. By defining $f_k(\theta) \triangleq (\beta - \theta K)\log_2(1 + \mathrm{E}\{\gamma_k\})$, the optimization problem (43) can be rewritten as

$$\max_{\theta} \sum_{i=1}^{K} f_k(\theta). \tag{53}$$

After some manipulations, the first and second derivatives of $f_k(\theta)$ can be expressed as

$$\frac{df_k(\theta)}{d\theta} = -K\log_2(1 + \mathrm{E}\{\gamma_k\}) + \frac{\beta - \theta K}{1 + \mathrm{E}\{\gamma_k\}}\frac{d\mathrm{E}\{\gamma_k\}}{d\theta}$$

$$\frac{d^2 f_k(\theta)}{d\theta^2} = -K\frac{1}{1 + \mathrm{E}\{\gamma_k\}}\frac{d\mathrm{E}\{\gamma_k\}}{d\theta} + \frac{-K}{1 + \mathrm{E}\{\gamma_k\}}\frac{d\mathrm{E}\{\gamma_k\}}{d\theta} + \frac{\beta - \theta K}{1 + \mathrm{E}\{\gamma_k\}}\frac{d^2\mathrm{E}\{\gamma_k\}}{d\theta^2} \tag{54}$$

$$= \frac{-1}{1 + \mathrm{E}\{\gamma_k\}}\left[2K\frac{d\mathrm{E}\{\gamma_k\}}{d\theta} + (\beta - \theta K)\frac{d^2\mathrm{E}\{\gamma_k\}}{d\theta^2}\right].$$

It follows that

$$\frac{d\mathrm{E}\{\gamma_k\}}{d\theta} = \frac{\bar{q}}{\sigma_d^2}, \quad \frac{d^2\mathrm{E}\{\gamma_k\}}{d\theta^2} = 0.$$

By substituting $\frac{d\mathrm{E}\{\gamma_k\}}{d\theta}$ and $\frac{d^2\mathrm{E}\{\gamma_k\}}{d\theta^2}$ into (54), one can obtain $\frac{d^2 f_k(\theta)}{d\theta^2} < 0$ in the desired region $\theta K \leq \beta$. Hence (53) is strictly concave function and the optimal $\theta$ can be obtained using bisection approach from $\sum_{k=1}^{K} \frac{df_k(\theta)}{d\theta} = 0$ [49].

Next we examine the optimal $\theta$ when $\mathrm{E}\{\gamma_k\} \gg 1$ (at high SNR regions). For these regions, the optimal $\theta$ will satisfy

$$\sum_{k=1}^{K} \left( -K \log_2(1+\mathrm{E}\{\gamma_k\}) + \frac{\beta - \theta K}{1+\mathrm{E}\{\gamma_k\}} \frac{d\mathrm{E}\{\gamma_k\}}{d\theta} \right) = 0 \Rightarrow -K \log_2(\mathrm{E}\{\gamma_k\}) + \frac{\beta - \theta K}{\theta} = 0$$

$$\Rightarrow \frac{\beta}{\theta} - K \log_2(\theta) = K \left[ 1 + \log_2 \left( \frac{\bar{q}}{\sigma_d^2} \right) \right]. \quad (55)$$

## Appendix C

## Covariance Matrix Estimation

As we can see from Fig. 1, the uplink and downlink data, and uplink pilot transmissions are executed in each $T_c$. This shows that the BS can exploit the uplink pilot and decoded data symbols to obtain sufficient samples for computing the covariance matrix of each UE. From the uplink pilot, one can have $\tilde{\mathbf{h}}_k$ which is defined as (35). For the given $\tilde{\mathbf{h}}_k$, we suggest a simple least squares estimation approach to maintain the statistical behavior of $\mathbf{h}_k$ as

$$\tilde{\tilde{\mathbf{h}}}_{kt} = \bar{\mathbf{U}} (\bar{\mathbf{U}}^H \bar{\mathbf{U}})^{-1} \tilde{\mathbf{h}}_k = \bar{\mathbf{U}} \tilde{\mathbf{h}}_k. \quad (56)$$

The second equality is due to $\bar{\mathbf{U}}^H \bar{\mathbf{U}} = \mathbf{I}$. Also, during the uplink data transmission, we will have

$$\mathbf{Y}_d = \tilde{\mathbf{U}}^H (\mathbf{H}\mathbf{X}^H + \mathbf{N}) \quad (57)$$

where $\tilde{\mathbf{U}}^H$ is the analog unitary matrix in the uplink data transmission phase [23] and it may vary from one coherence time to the other, and $\mathbf{X} = [\mathbf{x}_1, \mathbf{x}_2, \cdots, \mathbf{x}_K]$ are the transmitted symbols of all UEs with $\mathbf{x}_k \in \mathcal{C}^{(\beta-\alpha)\times 1}$. In practice, as each UE transmits its own data symbols independently, $\mathbf{x}_k$ are likely to be uncorrelated. For this reason, the term that contains $\mathbf{h}_k$ can be obtained from

$$\tilde{\tilde{\mathbf{h}}}_{kd} = \frac{1}{|\hat{\mathbf{x}}_k|} \tilde{\mathbf{U}} (\tilde{\mathbf{U}}^H \tilde{\mathbf{U}})^{-1} \mathbf{Y}_d \hat{\mathbf{x}}_k \approx \tilde{\mathbf{U}} \tilde{\mathbf{U}}^H (\mathbf{h}_k + \mathbf{N} \hat{\mathbf{x}}_k) \quad (58)$$

where $\hat{\mathbf{x}}_k$ are the decoded data symbols of the $k$th UE. Note that when $\beta - \alpha$ is higher, the second approximation term of this expression is closer to equality.

From the uplink pilot and data transmission phases of $L$ frames, one can compute the following sample covariance matrix for the $k$th UE

$$\tilde{\mathbf{R}}_k = \frac{1}{2L}(\tilde{\mathbf{R}}_{kt} + \tilde{\mathbf{R}}_{kd}) \tag{59}$$

where $\tilde{\mathbf{R}}_{kt} = \sum_{i=1}^{L} \tilde{\tilde{\mathbf{h}}}_{kit}\tilde{\tilde{\mathbf{h}}}_{kit}^H$, $\tilde{\mathbf{R}}_{kd} = \sum_{i=1}^{L} \tilde{\tilde{\mathbf{h}}}_{kid}\tilde{\tilde{\mathbf{h}}}_{kid}^H$, and $\tilde{\tilde{\mathbf{h}}}_{kit}$ and $\tilde{\tilde{\mathbf{h}}}_{kid}$ are the $\tilde{\tilde{\mathbf{h}}}_k$ obtained from (56) and (58) for each $T_c$ duration, respectively. It is clearly seen that $\tilde{\mathbf{R}}_k$ contains both the channel and noise information. As the noise is assumed to be white, it has an impact on the eigenvalues of $\tilde{\mathbf{R}}_k$ only. Thus, one approach of estimating the channel covariance matrix is

$$\hat{\mathbf{R}}_k = \mathbf{V}_t \boldsymbol{\Gamma}_{t+} \mathbf{V}_t^H + \mathbf{V}_d \boldsymbol{\Gamma}_{d+} \mathbf{V}_d^H \tag{60}$$

where $\boldsymbol{\Gamma}_t(\boldsymbol{\Gamma}_d)$ and $\mathbf{V}_t(\mathbf{V}_d)$ are the eigenvalue and eigenvector matrices of $\tilde{\mathbf{R}}_{kt}(\tilde{\mathbf{R}}_{kd})$, respectively, $\boldsymbol{\Gamma}_{d+} = [\boldsymbol{\Gamma}_d - \sigma_d^2 \mathbf{I}]_+$ and $\boldsymbol{\Gamma}_{t+} = [\boldsymbol{\Gamma}_t - \sigma_t^2 \mathbf{I}]_+$.

We would like to mention here that the current paper proposes a particular approach to estimate $\mathbf{R}_k$. And, as such an estimate is not necessarily optimal, better estimation accuracy can be achieved by considering different approach (for instance by incorporating weights in (59)). Furthermore, the estimation of $\epsilon_k$ from $\hat{\mathbf{R}}_k$ is still an open research topic and is left for future research.

<h2 style="text-align:center">REFERENCES</h2>